\documentclass[aps,prd,preprint,tightenlines]{revtex4}
\usepackage{graphicx,amssymb,mathrsfs,epstopdf}
\usepackage{placeins}  
\usepackage{amsmath}
\usepackage{verbatim}
\usepackage{url}
\usepackage{xspace}
\usepackage{hyperref}
\usepackage{subfigure}
\usepackage{fancyvrb}
\usepackage{upquote}
\usepackage{varwidth}
\usepackage{float}
\usepackage{rotating}
\usepackage{multirow}

\usepackage{fancyvrb}
\usepackage{color}

\DefineVerbatimEnvironment{code}{Verbatim}{fontsize=\small, 
commandchars=\\\{\}}
\DefineVerbatimEnvironment{example}{Verbatim}{fontsize=\small, commandchars=\\\{\}}

\def\mr{$M_{\rm R}$}
\def\r2{$R^2$}

\begin{document}

\title{Optimizing Event Selection with the Random Grid Search}

\author{P.\,C.~Bhat$^{\,a}$\footnote{pushpa@fnal.gov}, H.\,B.~Prosper$^{\,b}$\footnote{harry@hep.fsu.edu}, S.~Sekmen$^{\,c}$\footnote{ssekmen@cern.ch}, C.~Stewart$^{\,d}$\footnote{stewart@broadinstitute.org}}
\affiliation{
$^{a}$~Fermi National Accelerator Laboratory, Batavia, Illinois, 60510-5011, USA\\
$^{b}$~Department of Physics, Florida State University, Tallahassee, Florida 32306, USA\\
$^{c}$~Kyungpook National University, Daegu, South Korea\\
$^{d}$~Broad Institute, Boston, USA}

\begin{abstract}
The random grid search (RGS)  is a simple, but efficient, stochastic algorithm to find optimal cuts that was developed in the context of the search for
the top quark at Fermilab in the mid-1990s. The algorithm, and associated code, have been enhanced recently  with the introduction of two new cut types, one of which has been successfully used in searches for supersymmetry at the Large Hadron Collider. The RGS optimization algorithm is described along
with the recent developments, which are illustrated with two examples from particle physics. One explores
the optimization of the selection of vector boson fusion events in the four-lepton decay mode of
the Higgs boson and the other optimizes SUSY searches using boosted objects and the razor
variables. \\
\end{abstract}

\date{\today}

\maketitle









\tableofcontents

\section{Introduction}
\label{sec:introduction}


One of the primary tasks in data analysis is to isolate a potential signal from the background,
that is, from the noise,  by applying thresholds to one or more discriminating variables.  In 
particle physics, thresholds on discriminating variables are usually referred to as cuts, therefore, for ease of exposition we shall use this terminology
throughout this paper. 

In many fields, enormous resources  are needed to acquire the rapidly growing datasets. It is therefore not surprising that scientists and engineers have devoted considerable effort to develop 
highly effective methods 
to extract signals from data. 
In this paper, we describe a simple and efficient method for finding
optimal cuts, called the \emph{random grid search} (RGS), and recent generalizations of it.

Many powerful methods exist to discriminate between different 
classes of objects that combine multiple variables into a single discriminating function; for example, 
boosted decision trees and neural networks~\cite{Bhat:2010zz, Bishop}. These methods have been used with great success in several high-profile particle physics analyses (see, for example, Refs.~\cite{Abbott:1998dc,Abazov:2009ii, 
Chatrchyan:2012xdj, Aad:2012tfa}). Such methods are also
used routinely in particle identification (see, for example, Ref.~\cite{Aad:2015ydr, Khachatryan:2015hwa, Sirunyan:2017ezt, Aaij:2015yqa}). Recently, particle physicists have begun to explore the potential of deep neural networks~\cite{Baldi:2014kfa,Baldi:2014pta,Baldi:2016fql,Guest:2016iqz, Searcy:2015apa, Aurisano:2016jvx, Acciarri:2016ryt}.  Several tools that implement these methods are readily available for use in high energy physics applications, such as TMVA~\cite{Hocker:2007ht}, NeuroBayes~\cite{Feindt:2006pm}, Scikit-learn~\cite{scikit-learn}, Keras~\cite{chollet2015keras}, TensorFlow~\cite{tensorflow2015-whitepaper, 45381} and Theano~\cite{2016arXiv160502688short}.

However, even when more sophisticated methods are used, it is often useful to select a signal-enhanced
dataset by applying optimal cuts to one or more discriminating variables, which could
themselves be the outputs of sophisticated multivariate discriminants.  It is also 
demonstrably easier to discern the reason why a particular data sample was selected
if the decision is based on the application of cuts, especially if the cuts are to variables with
clear physical meaning.  
Decision trees gained favor in particle physics precisely because the decisions they furnish are manifest in the cuts they provide.
Note, however, that while this is true for a single decision tree, it is not so for the boosted variety, which
is an average over many  trees.   Perhaps the most compelling reason to consider methods based
on cuts is that
many analyses in particle physics are still cut-based. Therefore, simple efficient methods  for
finding optimal cuts, such as the random grid search, are still of considerable interest. 

The principle behind the random grid search is simple: search for cuts where they are
most likely to be useful.  Since the goal is to find signal-enhanced regions, it seems reasonable that
a cut-based search algorithm
focus on signal-like regions rather than explore the entire phase space. Given a large ensemble of sets of cuts, the RGS algorithm applies each set of cuts to a collection of signal and background events and counts the number of signal and background events, $s$ and $b$, respectively, that pass the cuts. Using these counts, any measure of the quality of the
cuts that  can be calculated from them, typically a measure of the signal significance, can be used to find the optimal cuts. The key to the effectiveness of the algorithm is the use of importance sampling: the location of
the cuts is determined by the distribution of the signal. 

The implementation of the RGS algorithm
in the software package described in this paper runs sequentially. But, the algorithm can be trivially parallelized, for example by using the CERN package {\tt PROOF}~\cite{Ballintijn:2006ni}. One
could, for example, apply a 
large number of cuts to multiple signal and background files in parallel.

The random grid search was
developed in the early 1990s by Bhat, Prosper and Stewart~\cite{rgs0,Amos:1995tn} during the successful search for the top quark at Fermilab~\cite{Abachi:1995iq, Abe:1995hr}. Subsequently, the method was used in several analyses by the D0 Collaboration, including the
search for first generation scalar and vector leptoquarks~\cite{Abazov:2001mx} and 
the measurement of the $t\bar{t}$ cross section~\cite{Abazov:2002gy}. The random grid search algorithm (though not the specific program described here) has been used by 
the CMS Collaboration in its search for $B^0_s \rightarrow \mu^+\mu^-$ and $B^0 \rightarrow \mu^+\mu^-$~\cite{Chatrchyan:2012rga}.  It was also used as a central element in~\cite{Strobbe:2011lta}, which devised a Bayesian method to optimize new physics searches over generic physics models with free parameters. Recently, the algorithm was significantly extended at CERN in the context of various searches for supersymmetry at the LHC using the razor kinematic variables~\cite{Rogan:2010kb}. Clearly, cut-based analyses are still important, which is the motivation for describing here the generalized version of the publicly available RGS package.

The paper is organized as follows. The random grid search algorithm is described in Section~\ref{sec:RGSalgorithm}. The utility of the algorithm is illustrated in Section~\ref{sec:examples} using a series of examples. The paper is summarized in Section~\ref{sec:summary}. For completeness, the Appendix contains
a detailed user manual for the RGS package.

\section{The Random Grid Search}
\label{sec:RGSalgorithm}



Suppose we wish to select a signal-enriched sample from a large dataset by cutting on two variables $x$ and $y$. The obvious way is to consider a grid of
$N^d$ points $(x_i, y_i), i = 1,\cdots, N$, with $d = 2$, quantify the quality of each 
\emph{cut-point}
$(x > x_i)$ \, and \, $(y > y_i)$ in separating signal from background, and select the best cut-point. Geometrically, a cut-point is the vertex defined by the lines aligned with the coordinate axes that meet at a point, here  $(x_i, y_i)$, while algebraically it is defined by expressions of the form $(x > x_i)$ \, \& \, $(y > y_i)$. 
The well-known
problem with this approach is the ``curse of dimensionality", the exponentially fast
rise in the number of cut-points as $N$ and $d$ increase. The curse is actually much worse than the numbers imply because the vast majority of 
cut-points are likely to be highly sub-optimal and therefore useless. Consequently, even if the resources were available to
scan billions of cut-points, most of these resources would be wasted.

\begin{figure}[htbp]
\begin{center}
\includegraphics[width=0.48\linewidth]{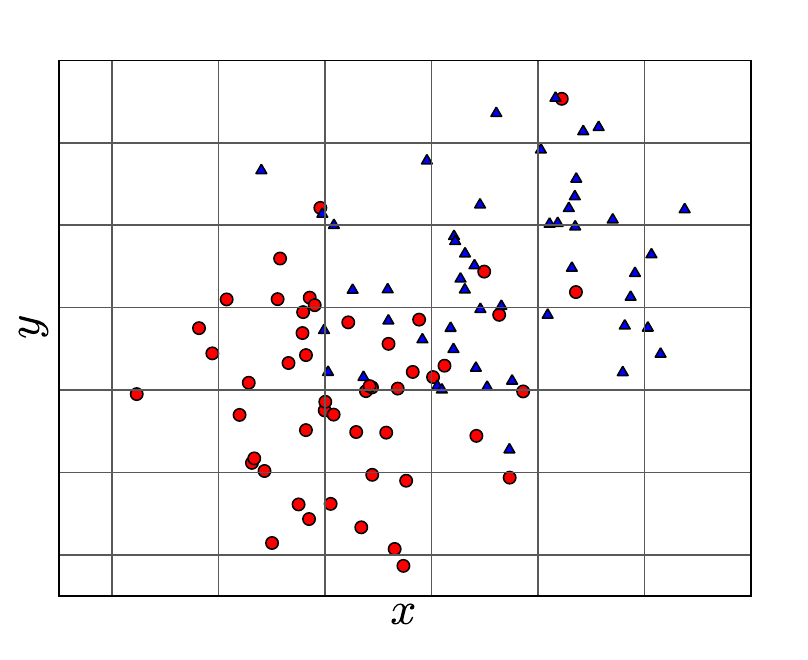}
\includegraphics[width=0.48\linewidth]{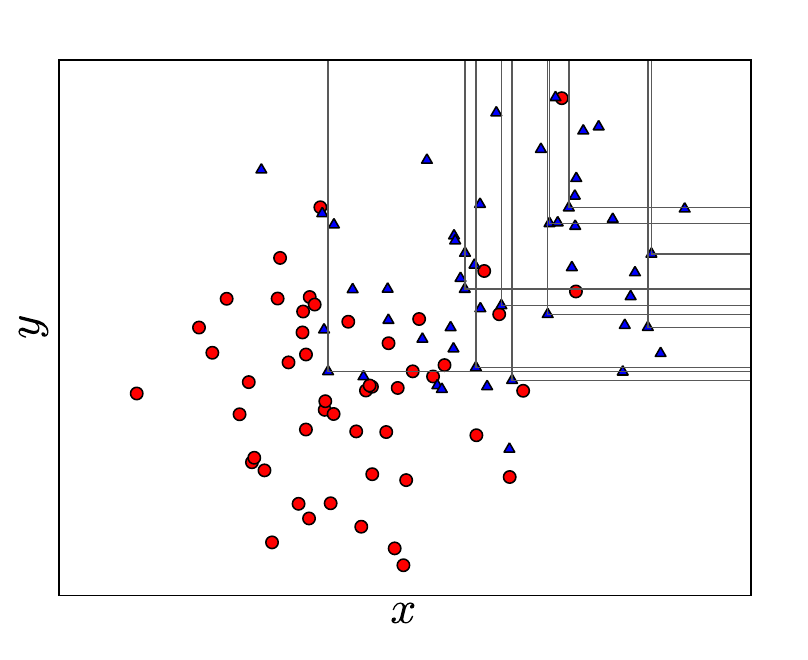}
\caption{A regular grid search (left) versus a random grid search (right) in a 2-dimensional space defined by the discriminating variables $x$ and $y$.  The triangles represent the signal events, the circles represent the background events and the lines show the values of $x$ and $y$ where the cuts are applied~\cite{Amos:1995tn}. Geometrically, a cut-point is the intersection of the lines.}
\label{fig:darkmatter}
\end{center}
\end{figure}

As alluded to, the key idea of the random grid search (RGS) is to use the \emph{signal distribution} as
the distribution of cut-points, a choice that focuses the search for good cuts to the region where they are more 
likely to be found.  By using the signal distribution, resources are focused most efficiently on cuts that best characterize the signal of interest.  
Note, however, that nothing in RGS precludes the use of distributions other
than the signal in order to define the cut-points.  For example, in the cases where signals are not exactly known, generic signal distributions can be used for characterizing the signal classes.
Since the distribution of the signal determines the cut-points, the lines drawn through them,
as illustrated in Figure~\ref{fig:darkmatter}, form
a random grid, hence the name of the algorithm.
Figure~\ref{fig:darkmatter} shows the difference between cuts on a regular grid in two dimensions (2-d)
and a random grid and  illustrates the potential efficiency gain in isolating signal-enriched regions using
the RGS algorithm.

The RGS algorithm is an effective way to 
optimize a signal's presence over backgrounds in order
 to discover or measure that signal.  Another useful application is finding the purest control regions in an analysis, where one needs to maximize the presence of a certain process, or a kinematic characteristic.

\subsection{Algorithm}
Cut-points are read from a file of events that can consist of signal events, points randomly sampled about the signal, events from a signal-like model, events for a process for which a control region is to be defined; indeed, the cut-points can be based on any collection of events.
The file from which cuts are read is referred to as the \emph{cut file}. The signal and background event files to which the cut-points are applied are referred  to as \emph{search files}. The type of cuts to be applied, which are described below, are specified in a \emph{cuts definition file}. Conceptually, the core algorithm proceeds as follows. (The syntax {\tt [A]} denotes a list of objects of type {\tt A}.)
\vbox{
\begin{verbatim}
	for cut-point in [cut-point]:
	       for search-file in [search-file]:
	              event-count = 0
	              for (weight, point) in search-file:
	                    passed = true
	                    for cut in cut-point:
	                           if cut fails: 
	                                 passed = false
	                                 break
	                    if passed:
	                           event-count = event-count + weight
	write out event-counts for each cut-point     
\end{verbatim}
}
There is an obvious place to parallelize this algorithm  using a {\tt PROOF}-like system designed to run over multiple files in parallel, namely, the loop over search files. For example, rather than apply cut-points to a search
file with $10^7$ events, one could execute one hundred jobs in parallel each running over 
$10^5$ events per search file. This requires
no change to the existing RGS package. And, of course, one could split cut files across multiple
jobs.

Associated with every (cut-point, search-file) pair is an {\tt event-count} variable  that accumulates the (weighted) sum of events that pass the cut-point. Therefore, every cut-point that is written out after the RGS algorithm has completed is associated with one or more event-count variables, one variable for each search-file.  The event counts can be used to compute various statistical measures of signal significance in order to find the best cut-point, that is, the best set of cuts.  
The calculation of signal significances is a separate step performed on the output of the RGS algorithm.  

A common measure of signal significance is (see, for example, the Statistics section in~\cite{Olive:2016xmw})
\begin{align}
Z  & = \sqrt{2 K(x)}, \nonumber \\
\textrm{where } K(x) & = b [ (1 + x) \ln (1 + x) - x], \quad\textrm{and}\quad x = s / b,
\label{eq:Z}
\end{align}
which reduces to $Z = s / \sqrt{b}$ when $x << 1$, that is, when the signal is very much
less than the background.  Equation~(\ref{eq:Z}) can be derived from several
perspectives. For example, if the data strongly favors the signal plus background
distribution over the background-only
distribution then it is likely that a signal is actually present. For discrete probability distributions, the Kullback-Leibler divergence~\cite{KL} 
\begin{align}
D(P||Q) & = \sum_n P(n) \ln \frac{P(n)}{Q(n)},
\end{align}
is a parameterization-invariant measure of the dissimilarity between two distributions $P$ and $Q$, which is
zero if, and only if, $P = Q$. If we let $P(n) = \textrm{Poisson}(n, s + b)$, that is, the signal plus background probability distribution for count $n$, and we let $Q(n) =  \textrm{Poisson}(n, b)$, the
background-only probability distribution, it is readily shown that $D(P || Q) = K(x)$.

\subsection{Cut types}
In the original RGS implementation in the mid-1990s~\cite{Amos:1995tn}, only \emph{one-sided}
cuts were implemented:
\begin{align}
				x &> x_i,  &\textrm{or}, \nonumber\\
				x &< x_i, 	&\textrm{or}, \nonumber\\
				|x| &> |x_i|, &\textrm{or}, \nonumber\\	
				|x| &< |x_i|,
				\label{eq:one-sided}			
\end{align}
From these, higher dimensional cuts (cut-points) are constructed through the logical AND of 
one-sided
cuts. For example, suppose we wish to select events based on the transverse momentum ($p_\textrm{T}$) and rapidity ($y$) of the highest $p_\textrm{T}$ jet in each event. Each cut-point might be defined by
\begin{align}
				(p_\textrm{T} > p_{\textrm{T}i}) \quad &\& \quad (|y| < |y_i|),
\end{align}
that is, by the AND of the two one-sided cuts. 

We have recently extended the RGS algorithm to permit the use of two other generic ways of imposing cuts: 
i) the \emph{two-sided cuts}, 
e.g., 
\begin{align}
	(p_\textrm{T} > p_{\textrm{T}i}) \quad & \& \quad (p_\textrm{T} < p_{\textrm{T}j}),
\end{align}
where the indices $i$ and $j$ denote different cut-points from the same cut file, 
and ii) the \emph{staircase cuts}, which consist of the OR of cut-points  for two or more variables, e.g.,
\begin{align}
	[(p_\textrm{T} > p_{\textrm{T}i}) \quad &\& \quad (|y| < |y_i|)] \quad\textrm{OR} \nonumber\\
	[(p_\textrm{T} > p_{\textrm{T}j}) \quad &\& \quad (|y| < |y_j|)] \quad\textrm{OR} \nonumber\\
			:  \quad\quad & \quad\quad :
\end{align}
In order to create either a two-sided or a staircase cut,  every cut-point in the cut file is associated with one or more randomly selected cut-points from the same file. One extra cut-point is needed to create a two-sided cut and typically several extra cut-points (defined by the number of steps in the staircase cut) are used to create a staircase cut. The current version of RGS
therefore implements three classes of cut, that is, \emph{cut types}, one-sided, two-sided, or staircase with the last two types formed from appropriate combinations of the cut types listed in Eq.~(\ref{eq:one-sided}). We provide detailed examples of the use of these cut types in Section~\ref{sec:examples} along with detailed syntax examples in the users manual in Appendix~\ref{sec:manual}.

\section{Physics examples}
\label{sec:examples}


As noted in the introduction, sophisticated multivariate methods of discrimination are now used
routinely in particle physics. However, cut-based methods still have a useful role to play. 
One motivation is that the publication of cut-based benchmark analyses is an important
service to the scientific community, especially to those who wish to reproduce published experimental results. The point is that,  even if, for excellent reasons, one publishes an experimental result 
that uses a state-of-the-art multivariate discrimination method, it is still of great value also to
publish an optimized cut-based analysis that can be more readily reproduced by those not involved
in the original analysis.  On the experimental side, algorithms such as RGS provide a straightforward way to improve analyses by improving the cuts they use. Indeed, given the quantity of data at our disposal, there is really no
good reason to dispense with cut optimization. Furthermore, optimized 
cut-based analyses provide benchmarks with respect to which the scientific utility of more sophisticated
methods can be judged. 

Since RGS is an algorithm for optimizing cuts, it can be applied to any set of variables. It may sometimes
be advantageous to apply RGS to functions of these variables rather than to the original ones, such as the linear function that de-correlates them. In this case, the implied cuts on the 
original variables will, in general, no longer be orthogonal. Thus the RGS algorithm is more
flexible than one might suppose despite its simplicity. 


In this section, we show two examples of the use of RGS to produce optimized cut-based
analyses.
We first illustrate the one-sided and two-sided cut types using a simplified analysis of 
$\textrm{H} \rightarrow \textrm{ZZ} \rightarrow 4\ell$ final states.  
Then we show how all three RGS cut types can be employed in exploratory analyses that may suggest 
unexpected ways of selecting data.



%
\subsection{Exploring $\textrm{H} \rightarrow \textrm{ZZ} \rightarrow 4\ell$ with RGS}
The discovery of the Higgs boson at CERN in 2012~\cite{Aad:2012tfa, Chatrchyan:2012xdj} marked  the completion of the Standard Model (SM)  and brought  to a close a fifty-year period of extraordinary
advances in particle physics. Almost immediately,  the Higgs boson went from a celebrated discovery  to 
a prized tool in the search for new physics. One area in particular is receiving increasing attention, namely, the production of the Higgs boson in the vector boson fusion (VBF) mode (see, for example, Refs.\,\cite{CMS:2016jjx,ATLAS:2016gld,Aaboud:2016cns,Cacciari:2015jma})  in
which weak vector bosons are radiated from initial state quarks and fuse to form a Higgs boson. New
physics could modify details of weak vector boson fusion events.

We consider simplified analyses of the processes
\begin{align}
p + p & \rightarrow H \rightarrow Z/\gamma +  Z/\gamma  + X, \\
p + p & \rightarrow Z/\gamma + Z/\gamma + X,
\end{align}
where the Z bosons decay to two charged leptons, either $ee$ or $\mu\mu$, thereby producing final states
with at least four charged leptons. The quantity $X$ denotes additional objects in the final state.
The dominant production mode of the Higgs boson is the gluon gluon fusion (ggF) process, while the 
dominant background arises from the production of two Z bosons without  
the intermediary Higgs boson, processes we shall refer to as the ZZ background. The principal
experimental signature of vector boson fusion processes is the appearance of at least two high rapidity jets among
the set of objects $X$. We consider one benchmark analysis, modeled on the CMS analysis of the 4-lepton
final state~\cite{Chatrchyan:2013mxa} and three RGS optimized analyses. 

\begin{figure}
\begin{center}
\includegraphics[width=0.32\textwidth]{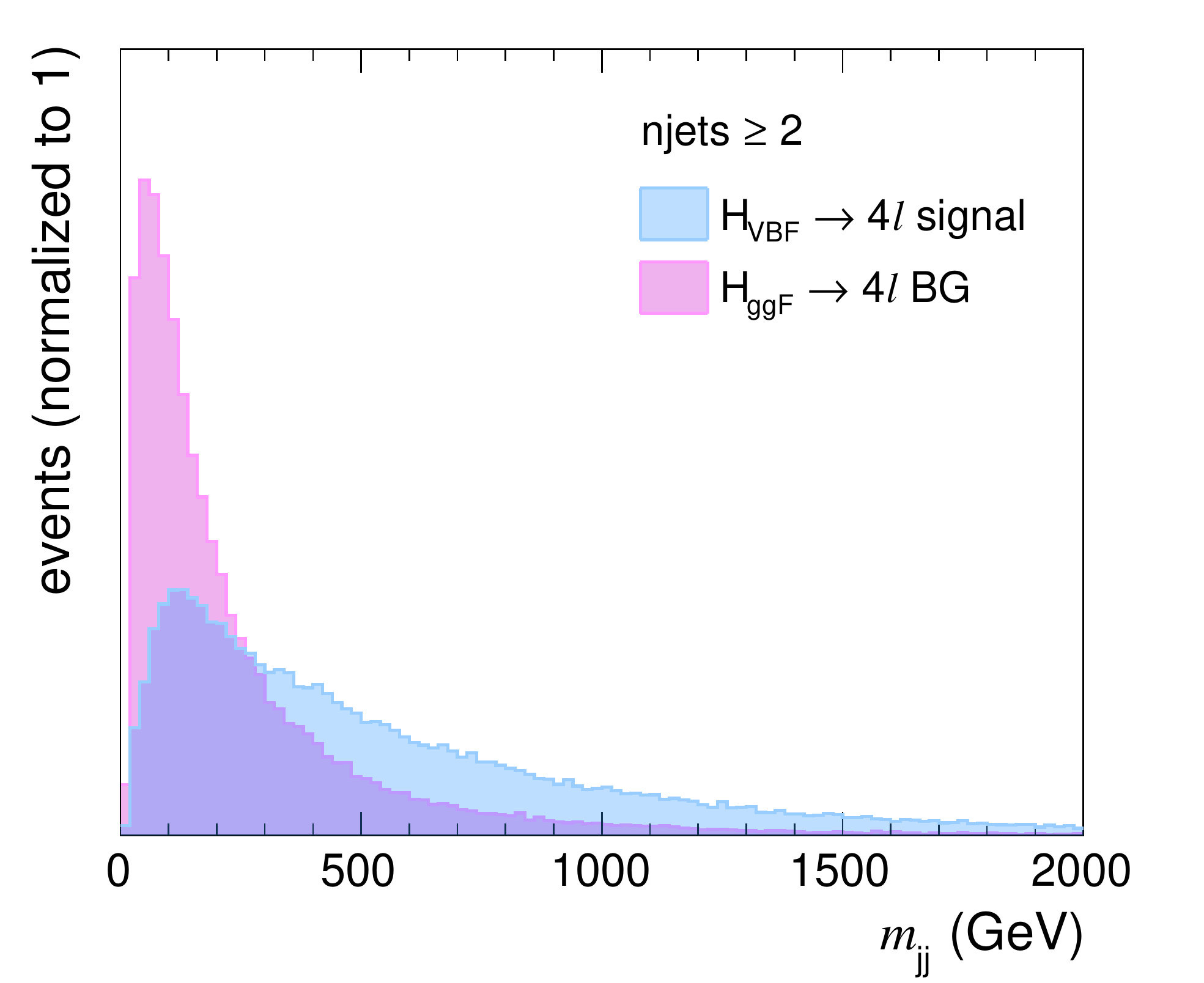}
\includegraphics[width=0.32\textwidth]{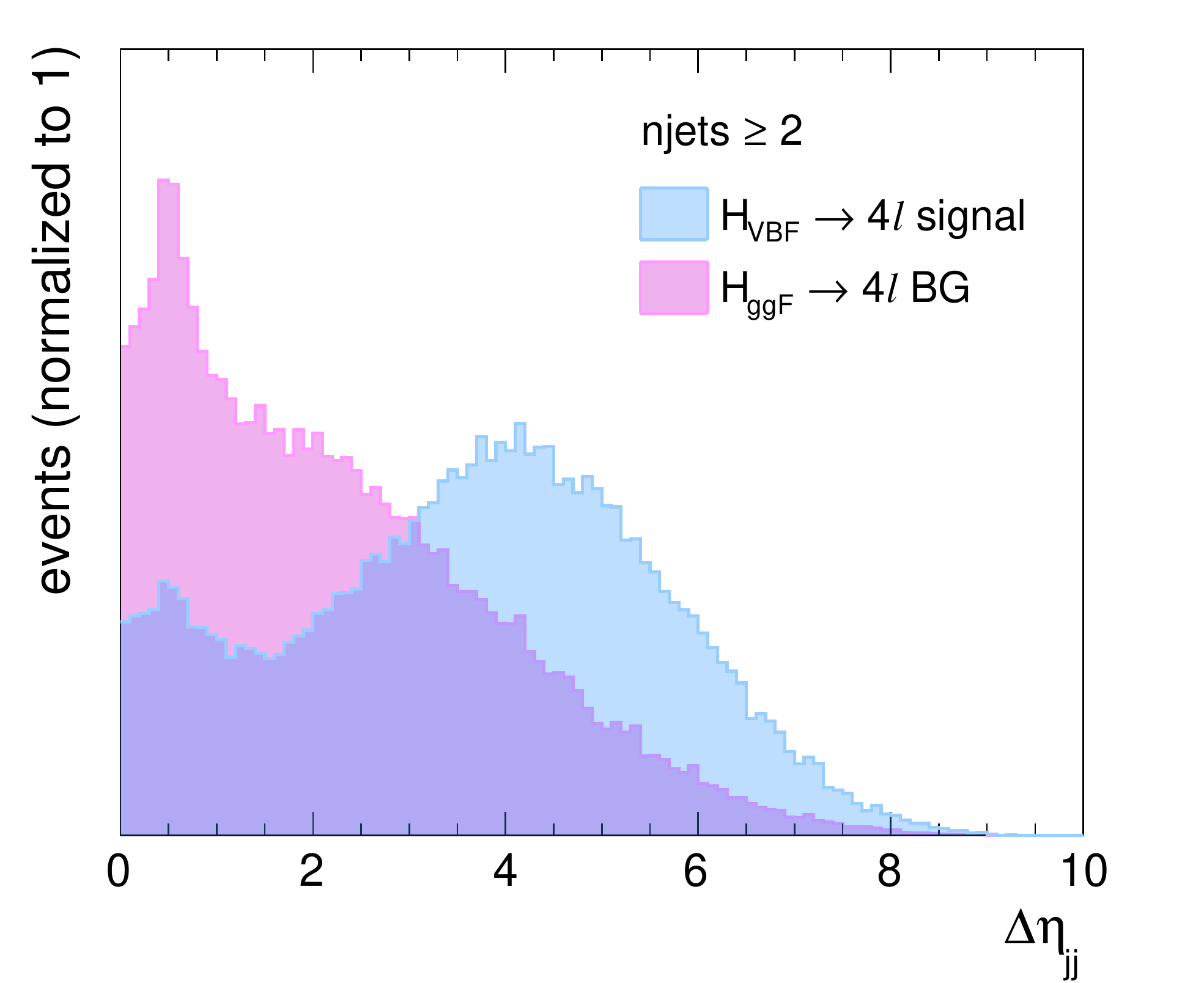} \\ 
\includegraphics[width=0.32\textwidth]{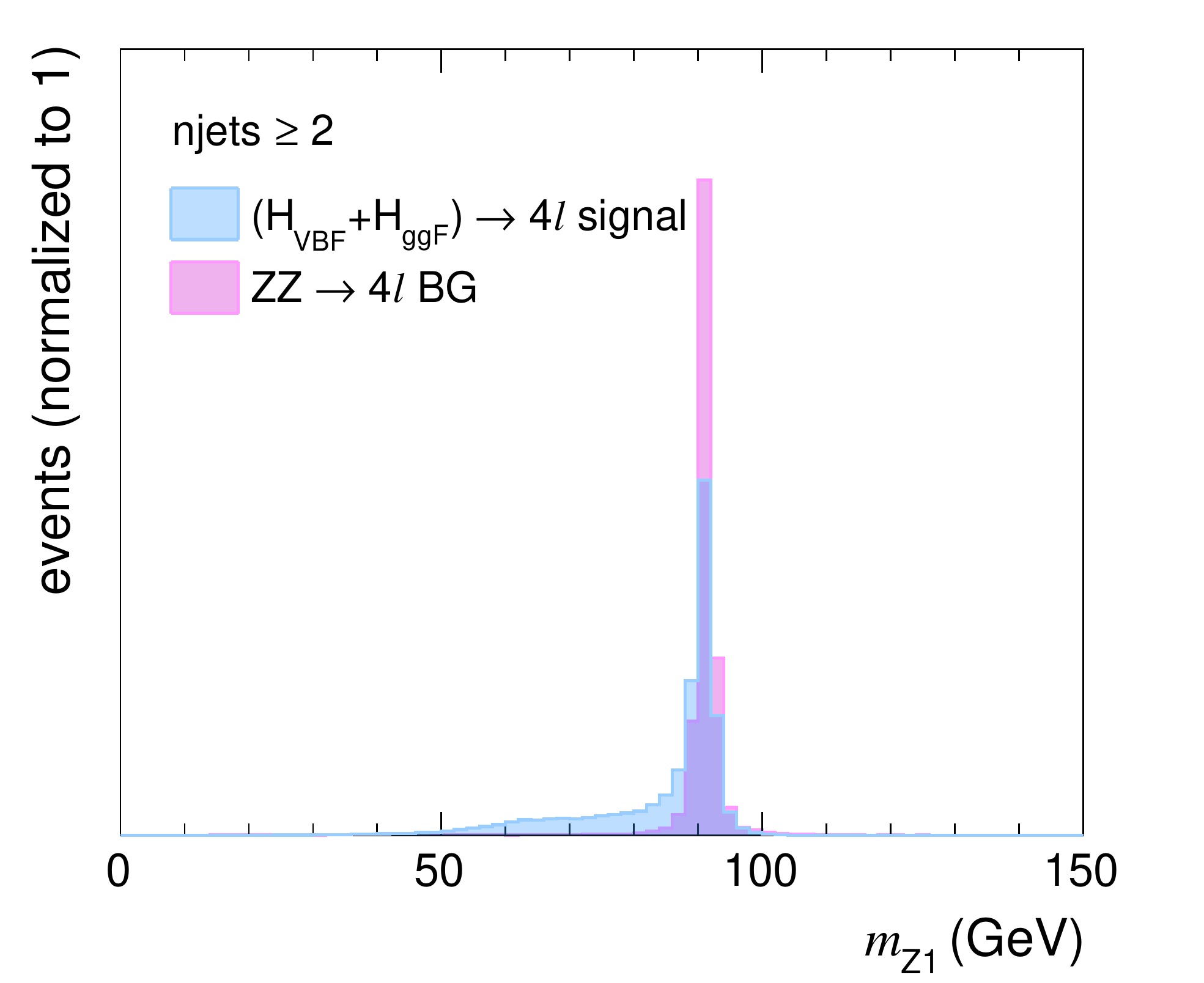}
\includegraphics[width=0.32\textwidth]{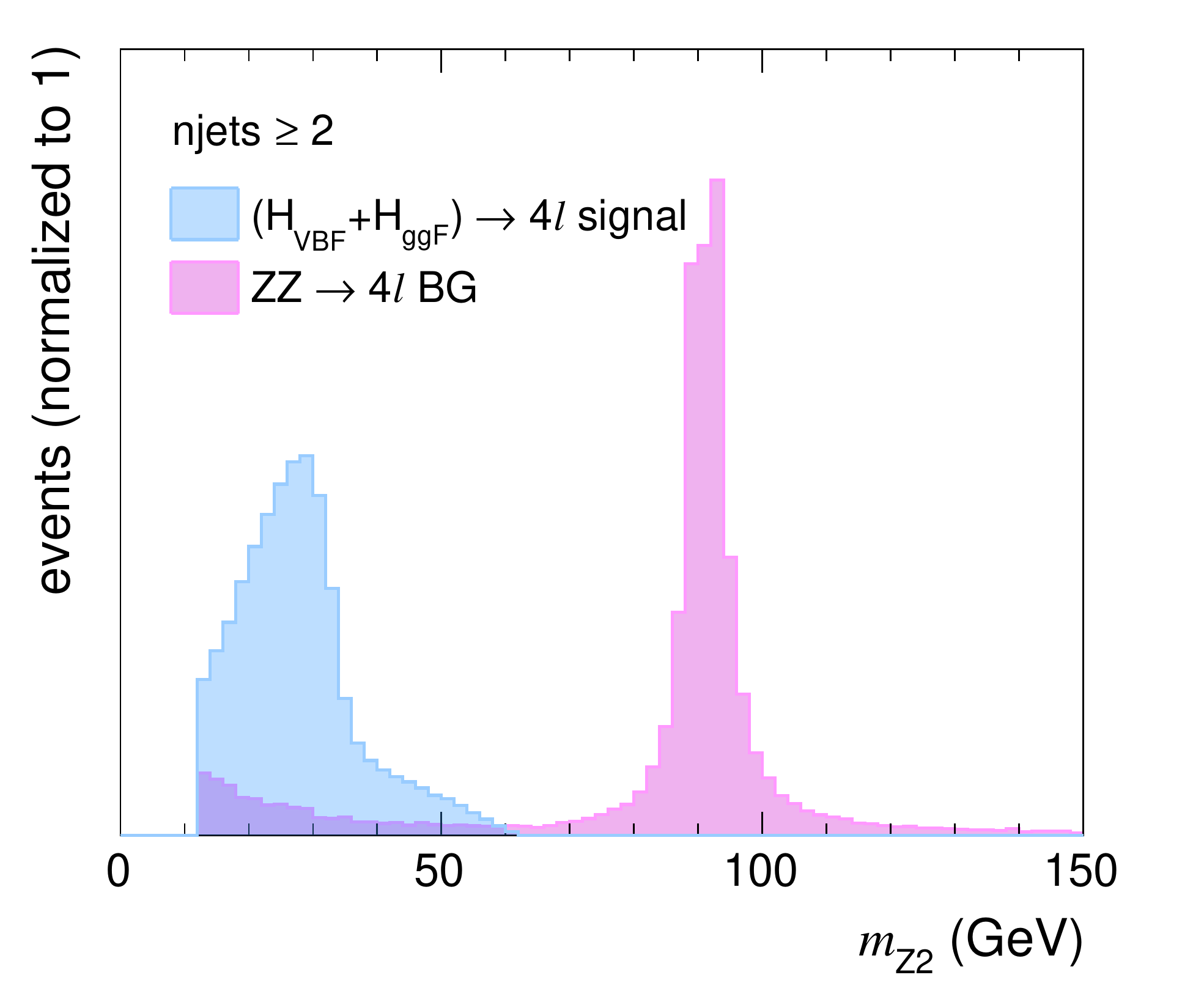} \\ 
\caption{(upper) Comparison of distributions of discriminating variables between the $\rm H_{VBF}$ signal and the $\rm H_{ggF}$ background processes (top row) and between Higgs boson events
and the ZZ background (bottom row). (top left) the di-jet mass $m_{jj}$, (top right) the absolute value of the di-jet rapidity difference $\Delta\eta_{jj}$, (bottom left) the mass distribution of the di-lepton
closest to the Z boson pole mass, which is referred to as Z$_1$, and (bottom right) the same for the second
di-lepton, which is referred to as Z$_2$.}
\label{fig:VBFggF}
\end{center}
\end{figure}

Higgs boson VBF processes are challenging because their cross sections
are smaller by an order of magnitude than those of the gluon gluon fusion (ggF) processes and they suffer from a paucity of distinctive features that distinguishes them from the ggF processes. Indeed,  to date, only three moderately effective VBF/ggF  discriminating observables  have been identified; they are the number of jets---specifically, the requirement that there be at least two jets, the mass $m_{jj}$ of the di-jet and $\Delta\eta_{jj}$, the absolute value of their rapidity difference. 
(Here, we neglect the jet masses and therefore do not distinguish between pseudo-rapidity, defined by $\eta = -2\ln\tan\theta/2$ where $\theta$ is the polar angle,
and the rapidity $y$.) The normalized distributions of the $m_{jj}$ and $\Delta\eta_{jj}$ observables, together with the mass distributions of the Z bosons are shown
in Figure~\ref{fig:VBFggF}. In our simplified analysis, we discriminate Higgs boson events from the ZZ background
using the masses, $m_{Z1}$ and $m_{Z2}$, of the two Z bosons, while the Higgs
VBF and ggF processes are discriminated using the di-jet observables.

The three classes of proton-proton collision events, yielding four charged leptons in the final state, VBF Higgs ($\textrm{H}_\textrm{VBF}$), ggF Higgs ($\textrm{H}_\textrm{ggF}$), and ZZ, were generated at the LHC center of mass energy of 13\,TeV using  {\tt PYTHIA8.209}~\cite{Sjostrand:2006za, Sjostrand:2007gs} with the Higgs boson mass set at 125\;GeV. We approximate the 
 response of the CMS detector with the public fast simulation package {\tt Delphes v3.2.0}~\cite{deFavereau:2013fsa} tuned to reflect the latest published CMS electron and muon identification efficiencies~\cite{Khachatryan:2015hwa, Chatrchyan:2013sba}. The default settings for {\tt PYTHIA8} were used, 
 together with a cut of $> 10\,\textrm{GeV}$ on the di-lepton masses ({\tt PYTHIA8} parameter
 {\tt 23:mMin}). In addition, we impose a minimum transverse momentum ($p_\textrm{T}$) cut of 5\,GeV ({\tt PYTHIA8} parameter {\tt PhaseSpace:pTHatMin}) on each of
 the products directly produced in the hard scatter. In particular, this is the minimum $p_\textrm{T}$
 of the Z bosons and of the forward quarks in the VBF processes. The cut is
 most relevant for the VBF processes and reduces the cross sections of the other processes by less than the quoted uncertainties in their predicted cross sections.

 The ZZ cross section at next-to-leading (NLO) accuracy is computed using 
 {\tt POWHEG-BOX-V2/ZZ}~\cite{Nason:2013ydw}, which for 
 a minimum $p_\textrm{T}$ cut on the di-lepton mass of  10\,GeV ({\tt POWHEG} parameter {\tt mllmin}) predicts a $p + p \rightarrow Z + Z \rightarrow e^+ + e^- + \mu+\mu^-$ cross section of $80.0 \pm 1.1\,\textrm{fb}$ and a $p + p \rightarrow Z + Z \rightarrow 4\ell, \,\, \ell = e, \mu$ cross section of $151.3 \pm 2.4\,\textrm{fb}$. The calculations were done using
 the {\tt CT14nlo}~\cite{Dulat:2015mca, Buckley:2014ana} set of parton distribution functions (PDF) and cross-checked with
{\tt MCFM v8.0}~\cite{Campbell:2015qma}, which for the same di-lepton mass cut and PDF set gives a
cross section of $89.7 \pm 1.1\,\textrm{fb}$  for the $ZZ \rightarrow e^+ + e^- + \mu+\mu^-$  process ({\tt MCFM} process code {\tt nproc 81}). 

The cross sections for the VBF and ggF Higgs boson processes in the $4\ell$ final states, with $\ell = e, \mu$, are taken from Ref.~\cite{deFlorian:2016spz}, which provides results computed
at different levels of accuracy; we use the cross sections labeled as NNLO + NLL QCD + NLO  EW. 
 For
the VBF processes, the cross section is
0.47\,fb, computed as follows: $\sigma(qq \rightarrow \textrm{H} + qq) \times \textrm{BR}(\textrm{H} \rightarrow \textrm{ZZ} \rightarrow 4\ell)$, where $\sigma(qq \rightarrow \textrm{H} + qq) = 3.782\,\textrm{pb}$ and $\textrm{BR}(\textrm{H} \rightarrow \textrm{ZZ} \rightarrow 4\ell) = 1.25\times10^{-4}, \,\,\ell = e, \mu$. Similarly, the ggF cross section is taken to be $\sigma(gg \rightarrow \textrm{H}) \times \textrm{BR}(\textrm{H} \rightarrow \textrm{ZZ} \rightarrow 4\ell) = 5.52\,\textrm{fb}$,
where $\sigma(gg \rightarrow \textrm{H}) = 44.14\,\textrm{pb}$. The calculations are for
a Higgs boson of mass 125\,GeV.
 
 We simulated $3 \times 10^5$, $10^6$, and $2 \times 10^6$ ggF, VBF, and ZZ events,
 respectively, and defined reconstructed leptons and jets with the following criteria.
 \begin{itemize}
 	\item {\bf Jets:} anti-$k_T$ jets with a jet size $R = 0.4$, where $R = \sqrt{\Delta\eta^2 + \Delta\phi^2}$ and $\eta$ and $\phi$ are the pseudo-rapidity and azimuthal
	angle, respectively. We required the jet $p_T > 20\,\textrm{GeV}$, and jet $|\eta| < 4.7$.
	\item {\bf Leptons ($\mathbf e$ and $\mathbf \mu$: } Generated leptons, to which the CMS lepton identification efficiencies are applied, with $p_T > 5\,\textrm{GeV}$ and $|\eta| < 2.4 $ for muons and $p_T > 7\,\textrm{GeV}$ and
	$|\eta| < 2.5$ for electrons. 
 \end{itemize} 
Events with at least two jets and four charged leptons are selected. The di-lepton with mass closest to the Z pole mass is labeled
 $\textrm{Z}_1$, while the other di-lepton is labeled $\textrm{Z}_2$.  Both Z candidates are
 required to have a 
 mass above 12\,GeV. The relative efficiency of the event selection criteria in this simplified
 analysis was found to be 12.4\%, 3.5\%, and 1.8\% for ggF, VBF, and ZZ, respectively.

Our benchmark analysis, with respect to which we can compare the results of the different RGS
optimizations, imposes the additional cuts, 
 \begin{itemize}
 	\item leading lepton $p_T > 20\,\textrm{GeV}$;
	\item next to leading lepton $p_T > 10\,\textrm{GeV}$;
	\item $40 < m_{\textrm{Z}1} <  120\,\textrm{GeV}$, and
	\item $12 < m_{\textrm{Z}2} <  120\,\textrm{GeV}$,
 \end{itemize} 
 which are the most important ones in the CMS Higgs to four-lepton analysis\,\cite{Chatrchyan:2013mxa}.
Furthermore, we required events to lie within the loose Higgs boson signal
 region defined by $100 < m_{4\ell} < 150\,\textrm{GeV}$, where $m_{4\ell}$ is the 4-lepton mass. In this region, the Higgs boson signal is a narrow 
 peak at 125\,GeV, while the ZZ background is relatively flat.  
The results of the full selection (labeled CMS) are shown in Table\,\ref{tab:Higgsopts} for an integrated luminosity of $100\,\textrm{fb}^{-1}$. This analysis of the three simulated event samples yields a Z significance of 1.03.  Clearly,
that number leaves considerable room for improvement, which is the goal of the RGS optimizations
described next. In this paper, we 
use the significance measure defined in Equation~(\ref{eq:Z}).

\begin{table}
\caption{Summary of the different Higgs optimizations (HOs) and results for the $H \rightarrow ZZ$ signals for
and integrated luminosity of $100\,\textrm{fb}^{-1}$. All results are for events that lie in the
Higgs boson signal region (see text), except for HO1, which uses events from the
full phase space of the simulated events. $Z_{max}$ is the maximum value of the Z significance.
We use the significance measure defined in Eq.~(\ref{eq:Z}).}
\begin{center}
\begin{tabular}{|l|l|c|c|c|c|c|c|c|}
\hline
Opt & State & $m_{jj}$ & $|\Delta\eta_{jj}|$ & $m_{Z1}$ & $m_{Z2}$ & $N_S$ & $N_B$ & $Z_{max}$ \\
\hline
\hline
         &  &  &  &  &  & ${\rm H_{\rm VBF}}$ & ${\rm H_{\rm ggF}+ZZ}$ &  \\ 
CMS & Analysis cuts ($var$)  & -- & -- & 40 -- 120 & 12 -- 120 & 5.78 & 29.48 & 1.03 \\
\hline
\hline
HO1& Before opt. ($var$) &  &  &  &  &  &  &  \\
       & \multicolumn{1}{r|}{(no $m_{4l}$ cut)  \hspace{6pt} } & -- & -- & $>12$ & $>12$ & ${\rm H_{\rm VBF+ggF}}$ & ${\rm ZZ}$ &  \\ 
       & Optimized vars & $\times$ & $\times$ & $\checkmark$ & $\checkmark$ &  &  & \\
       & After opt. ($var$) & -- & -- & $< 92.3$ & $< 32.5$ & 16.41 & 16.53 & 3.55 \\
\hline
\hline
HO2& Before opt. ($var$) & -- & -- & $>12$ & $>12$ & ${\rm H_{\rm VBF+ggF}}$ & ${\rm ZZ}$ &  \\ 
       & Optimized vars & $\times$ & $\times$ & $\checkmark$ & $\checkmark$ &  &  & \\
       & After opt. ($var$) & -- & -- & $58.4 - 94.6$ & $ 17.0 - 52.1$ & 20.06 & 5.80 & 6.10 \\
\hline
\hline
HO3& Before opt. ($var$) & -- & -- & $58.4 - 94.6$ & $ 17.0 - 52.1$ & ${\rm H_{\rm VBF}}$ & ${\rm H_{\rm ggF}+ZZ}$ &  \\ 
       & Optimized vars & $\checkmark$ & $\checkmark$ & $\times$ & $\times$ &  &  & \\
       & After opt. ($var$) & $>396.0$ & $>2.99$ & $58.4 - 94.6$ & $ 17.0 - 52.1$ & 2.38 & 2.79 & 1.27 \\
\hline
\hline
\end{tabular}
\end{center}
\label{tab:Higgsopts}
\end{table}%

\begin{figure}
\begin{center}
\includegraphics[width=0.49\textwidth]{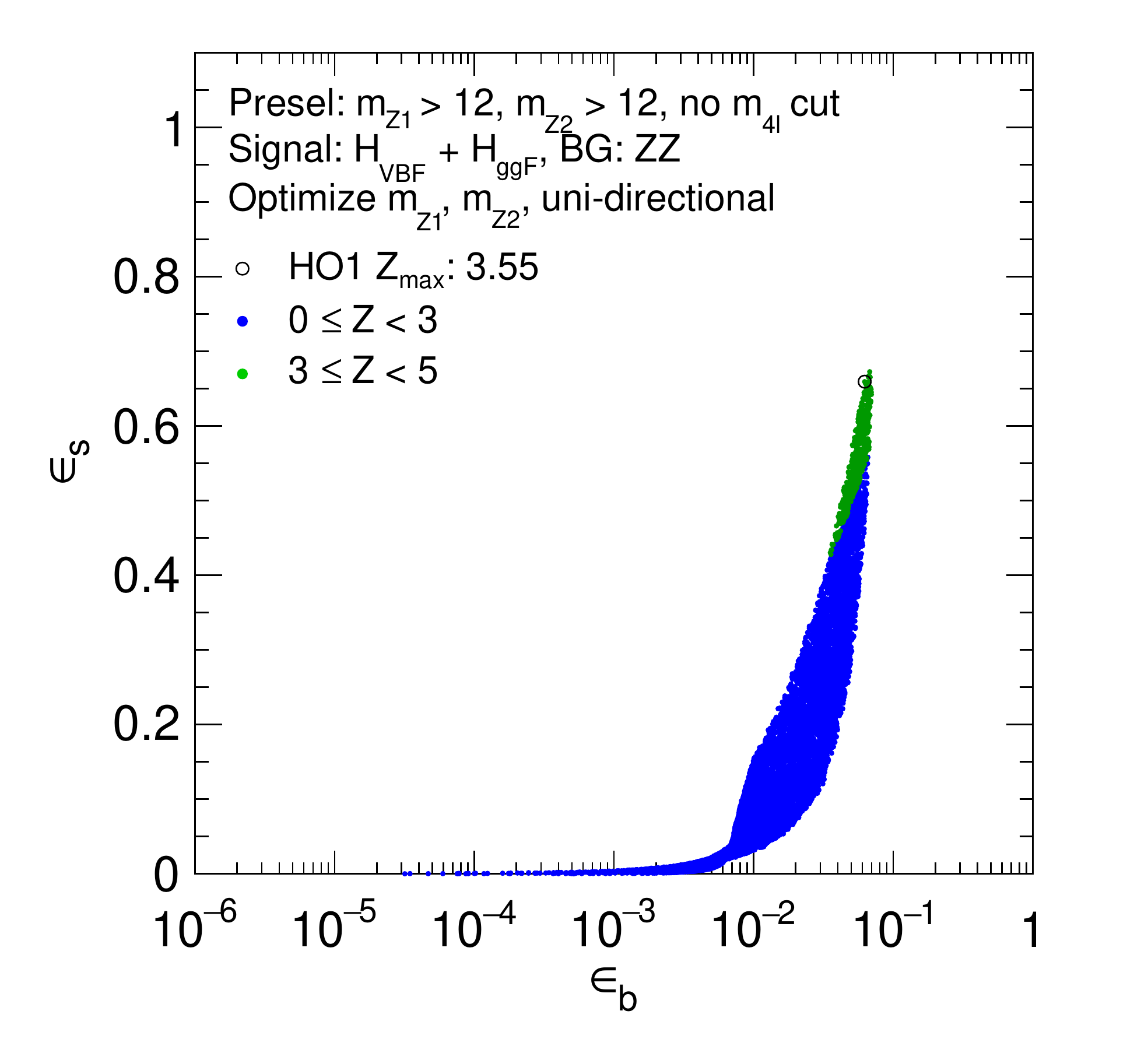}
\includegraphics[width=0.49\textwidth]{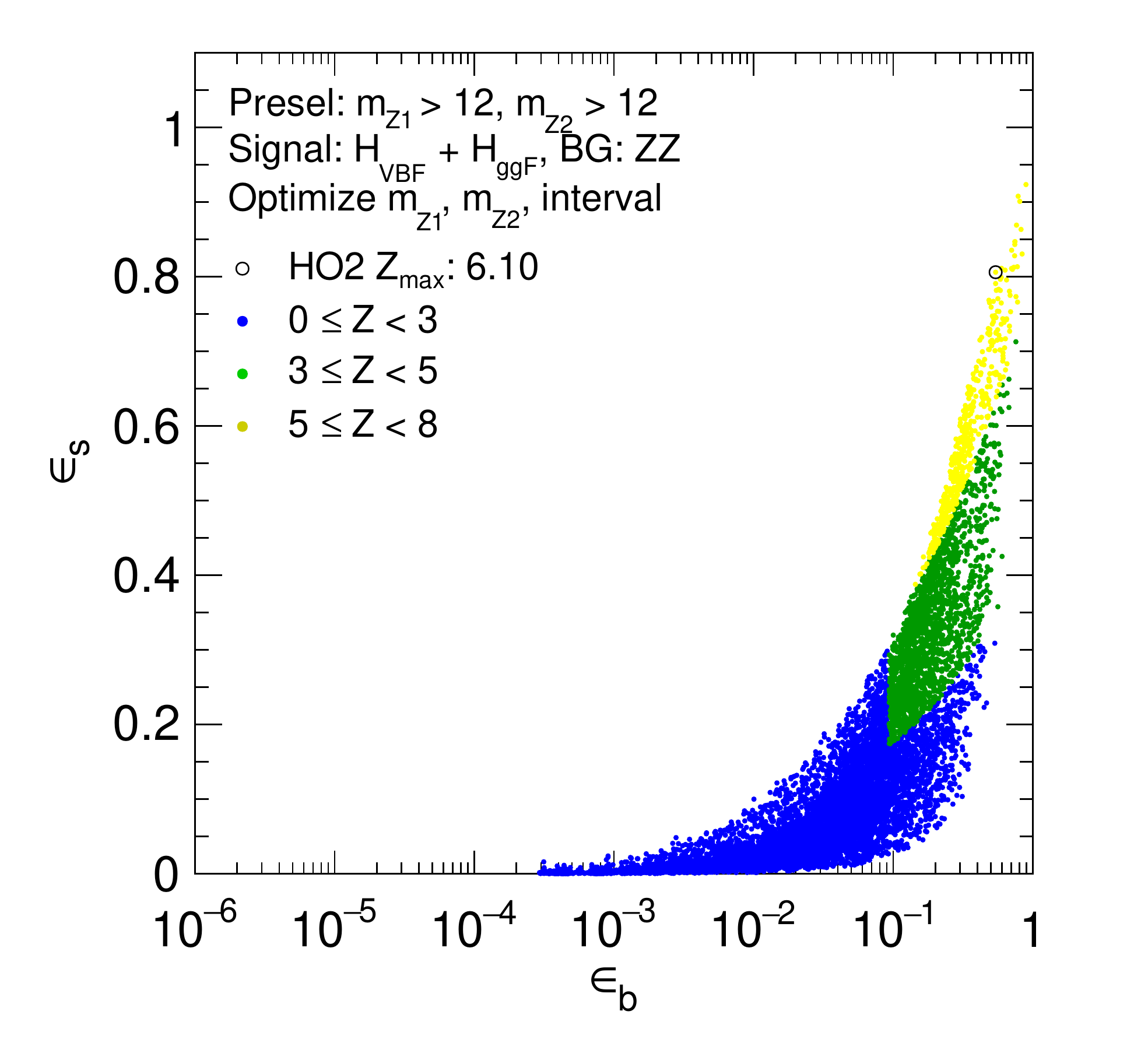} 
\includegraphics[width=0.49\textwidth]{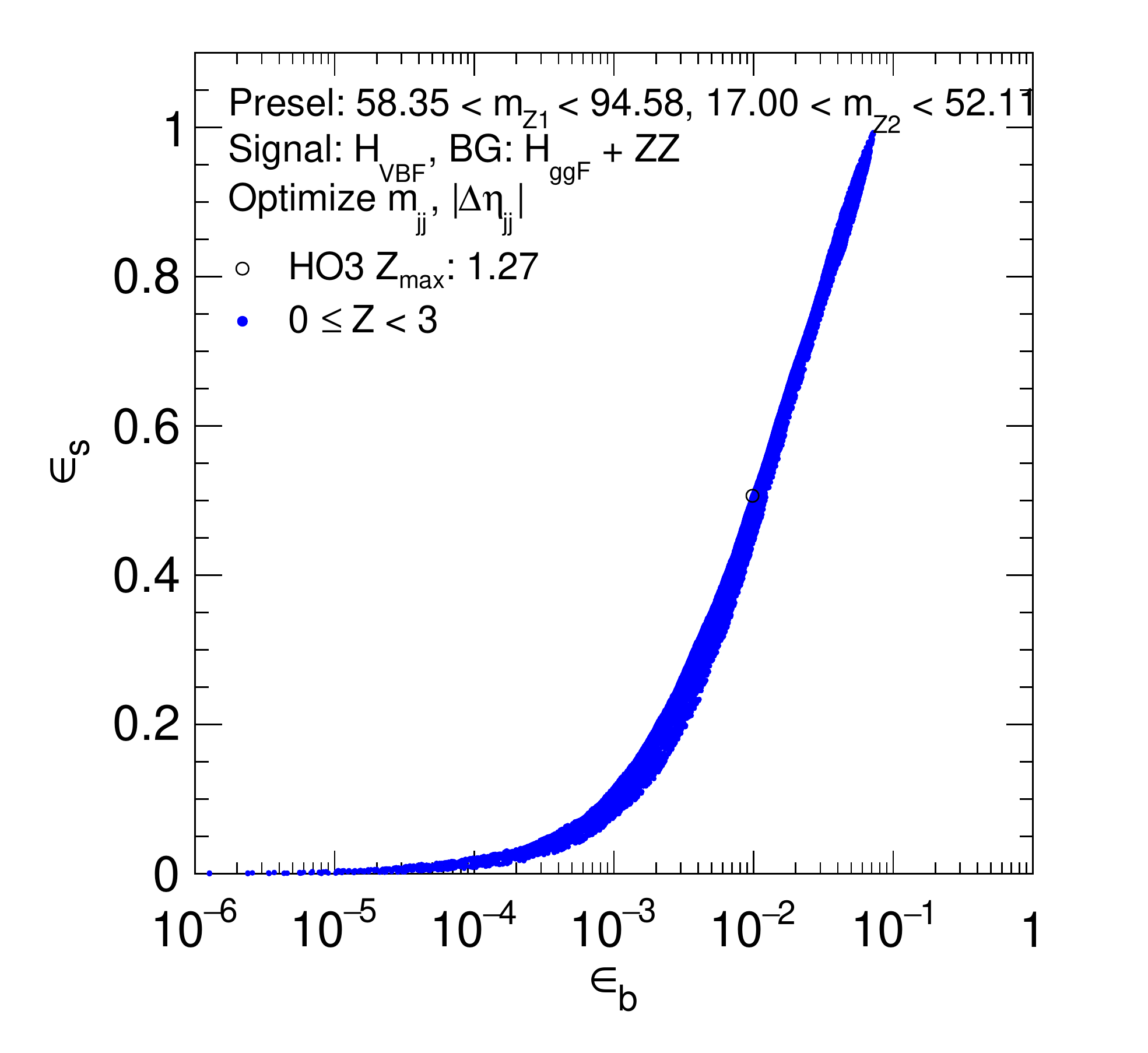} \\ 
\caption{Signal efficiency $\epsilon_s$ versus background efficiency $\epsilon_b$, and Z significance ranges for the three different classes of optimization in this study: HO1 (top left), HO2 (top right), HO3 (bottom).  For further details, see Table~\ref{tab:Higgsopts}.}
\label{fig:Higgsopts}
\end{center}
\end{figure}

\begin{figure}
\begin{center}
\includegraphics[width=0.49\textwidth]{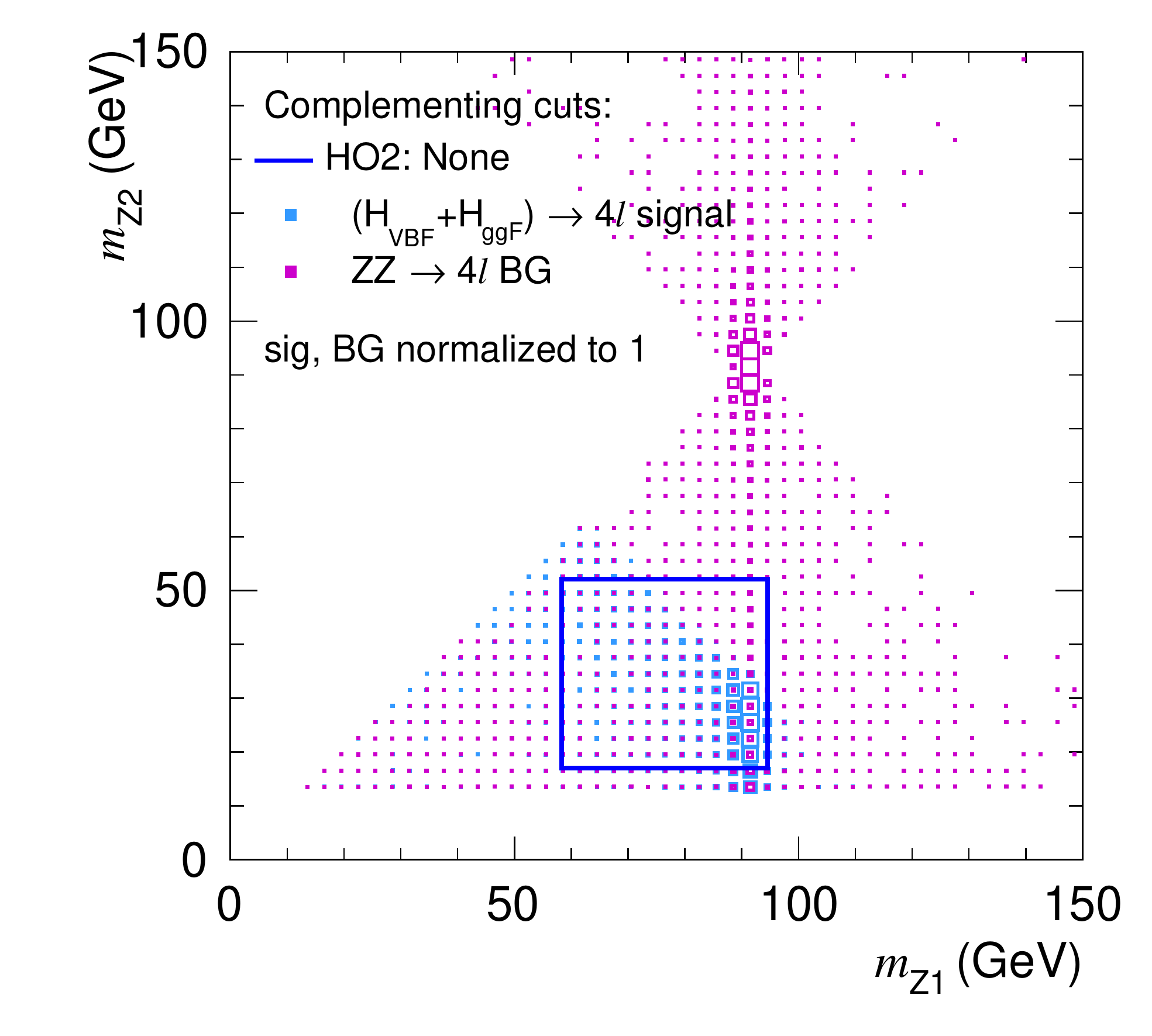}
\includegraphics[width=0.49\textwidth]{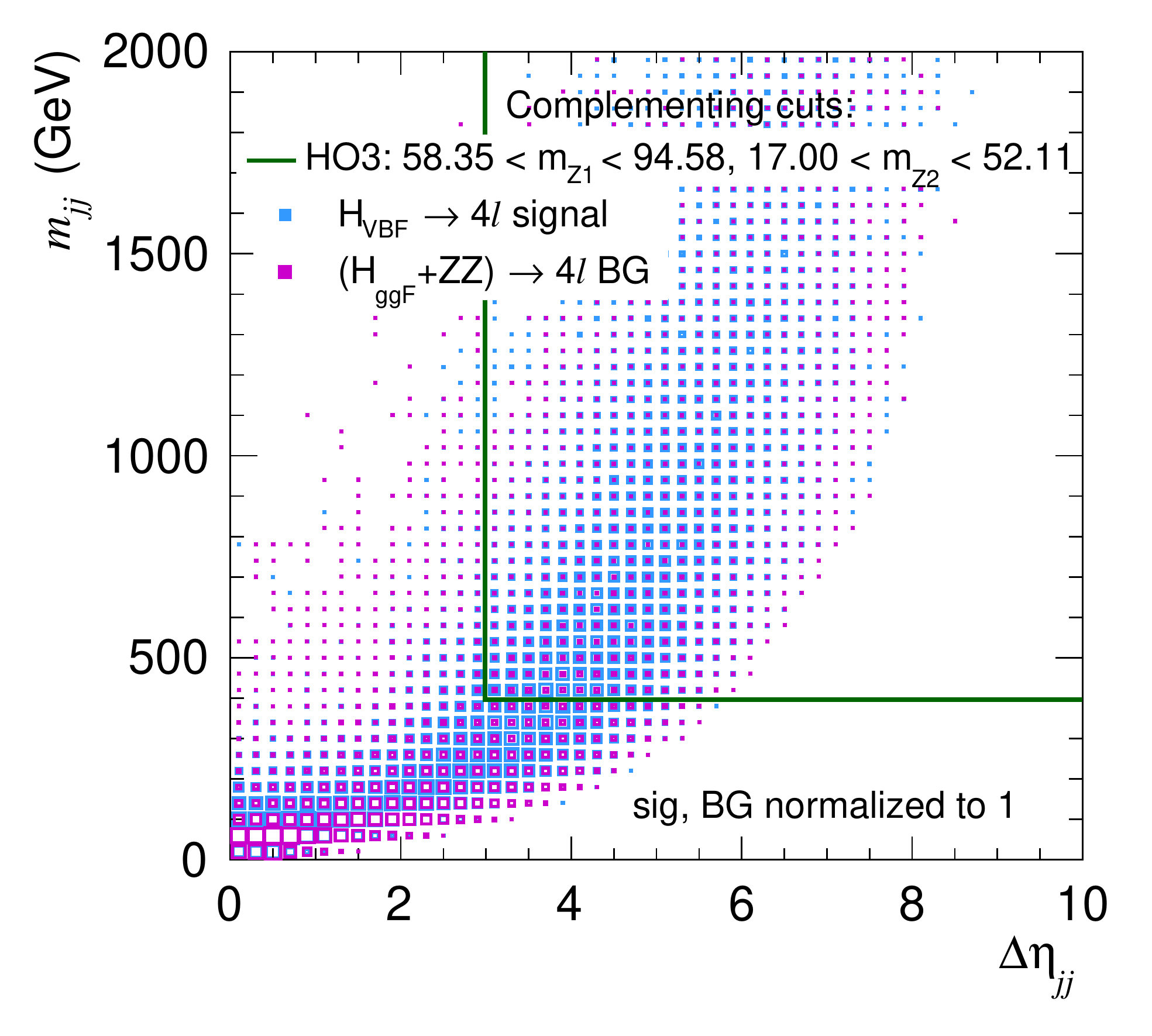}
\caption{The $m_{Z1} - m_{Z2}$  boundaries determined by the two-sided (i.e., box) cuts 
obtained using the HO2 optimization (left). The $|\Delta\eta|_{jj} - m_{jj}$ cut boundaries determined 
by the staircase cut using the HO3 optimization (right). }
\label{fig:Higgsopt}
\end{center}
\end{figure}

Three RGS optimizations are considered, each with $10^4$ cut-points.  Our strategy is to first find a subspace in which the event yield of Higgs boson events is enhanced relative to ZZ (in Higgs optimization HO2), and then, within that subspace, enhance the yield of VBF relative to ggF plus the remaining ZZ events (in HO3).  
The purpose of HO1 is merely to illustrate the use of the simplest form of cut, namely, one-sided cuts. These cuts are applied in the ($m_\textrm{Z1}, m_\textrm{Z2}$) plane in order
 to separate ggF+VBF events from ZZ events.  Since Higgs boson events tend to populate the low
 $m_\textrm{Z2}$ region, the appropriate one-sided cuts are $m_\textrm{Z1} < \textrm{cut}_1$ and
 $m_\textrm{Z2} < \textrm{cut}_2$, where the values of the tuple $(\textrm{cut}_1$, $\textrm{cut}_2$)
 are determined by the distribution of ggF+VBF events in this plane.  The results
 of the HO1 optimization, in which the $m_{4\ell}$ cut is omitted,  are shown in Table\,\ref{tab:Higgsopts}. The top left plot
 of Figure\,\ref{fig:Higgsopts} shows the signal efficiency versus the background efficiency. The HO1 optimization yields a Z significance of 3.55. 
 
 The second optimization (HO2) searches for suitable two-sided
 cuts in the  ($m_\textrm{Z1}, m_\textrm{Z2}$) plane.  As is the case for HO1, the goal for HO2 is to discriminate between Higgs boson events and ZZ events, except that this time we require
 events to lie within the Higgs boson signal region, as defined above, and thereby  
 benefit from the greatly reduced ZZ background within that region.
 The HO2 results are shown in Table\,\ref{tab:Higgsopts}. The ggF+VBF signal
 efficiency versus the ZZ background efficiency is shown in the
 upper right plot of Figure\,\ref{fig:Higgsopts} for different ranges of Z significance, while the best
 two-sided cuts of all those considered within the ($m_\textrm{Z1}, m_\textrm{Z2}$) plane is shown in the left plot of Figure\,\ref{fig:Higgsopt}. This pair of two-sided cuts yields a Z significance of 6.1. 
 
 It comes as no surprise that if one modifies HO1 by including in its definition
 the two-sided cut on $m_{4\ell}$ that defines the signal region, the Z significance improves; in fact, to 4.84. What is notable is
 the fact that the optimal two-sided cuts in the ($m_\textrm{Z1}, m_\textrm{Z2}$) plane do considerably better. Indeed, for the modified HO1 optimization
 to achieve the same Z significance as HO2 would require the former to use 60\% more data.
 
The third optimization (HO3) seeks to enhance VBF relative to ggF and ZZ by finding the optimal one-sided cuts, $m_{jj} > \textrm{cut}_1$ and
 $\Delta\eta_{jj} > \textrm{cut}_2$, for events constrained to lie within the two-sided cuts found by HO2.
 The results
 of HO3  are shown in Table\,\ref{tab:Higgsopts}, while the lower plot 
 of Figure\,\ref{fig:Higgsopts} shows the signal efficiency versus the background efficiency and
 the right plot of Figure\,\ref{fig:Higgsopt} shows the optimal one-sided cuts.
  The HO3 optimization yields a Z significance of 1.27, which is to be compared with 1.03 for our version of the
 CMS analysis. If analyses are restricted to cut-based methods, there is clearly more work to be done before an unambiguous VBF signal emerges in 
 this channel. But, again, it is notable that 50\% more data would be needed for the ``CMS" analysis
 to match the results of HO3.

 We chose this optimization problem because it is particularly challenging for a cut-based analysis due to the large overlap between the distributions for Higgs boson ggF and VBF events. 
 It is already accepted that a precision measurement
 of the VBF cross section in the Higgs 4-lepton final state will require multivariate
 discriminants~\cite{Chatrchyan:2013mxa,Aad:2014eva} presumably constructed using machine learning
 methods. But, as noted in the introduction, cut-based analyses nevertheless remain useful as
 benchmarks that can be readily implemented. Moreover, the outputs of multivariate functions are
 merely sophisticated variables to which RGS can be applied if an
 optimal VBF region is needed.
 
 In the next section, we describe a more sophisticated use of RGS optimization.

\subsection{Optimizing SUSY razor boost with RGS}
The random grid search is particularly useful for efficiently optimizing searches for new physics. We demonstrate its power and versatility by applying RGS to a supersymmetry (SUSY) search published by the CMS Collaboration~\cite{Khachatryan:2016zcu}.  This is a search for new physics in the jets, $b$ jets and missing transverse energy ($E_T^{miss}$) final state that uses high momentum,  that is, ``boosted",  W bosons whose decay products are merged into a single ``fat" jet.  The  razor kinematic variables~\cite{Rogan:2010kb} are used to discriminate possible SUSY signals from the Standard Model (SM) backgrounds.  The analysis mainly targets gluino decays to top squark-top quark pairs.

To illustrate how RGS can be useful in designing searches, we selected a SUSY signal point with parameters that lead to a gluino with mass of 1355\,GeV, a light top squark with mass of 409\,GeV and a lightest neutralino with mass of 252\,GeV.  The remaining sparticles are heavy and inaccessible at the 13\,TeV LHC.  In this scenario, gluinos decay exclusively to $\tilde{t}_1 t$ pairs, where the 
momenta of the top quarks are high due to the large $m(\tilde{g}) - m(\tilde{t}_1)$ mass difference.  Subsequently, since the mass difference $m(\tilde{t}_1) - m(\tilde{\chi}^0_1)$ is smaller than $m(t)$, top squarks directly decay to $Wb\tilde{\chi}_1^0$.  This decay structure leads to multiple jets, multiple $b$ jets, and a significant fraction of events with boosted $W$ bosons coming from the top quarks produced directly in the gluino decays.  The dominant SM background for this signal is $t\bar{t}+$jets. For simplicity, we consider only this background process in this study.  For the signal point, we calculated the SUSY mass spectrum from the given SUSY parameter point using the {\tt SOFTSUSY v3.3.1} package~\cite{Allanach:2001kg}, and the sparticle decays using the {\tt SUSYHIT} package~\cite{Djouadi:2006bz} (with {\tt SDECAY 1.5} and {\tt HDECAY 3.4}).  For both the signal and background, we used {\tt PYTHIA8.209}~\cite{Sjostrand:2006za, Sjostrand:2007gs} to generate LHC events at 13\,TeV.  As was done for the Higgs boson study, the response of the CMS detector was simulated using {\tt Delphes v3.2.0}~\cite{deFavereau:2013fsa}.  We calculated the NLO cross section for the SUSY signal using the {\tt Prospino 2} package~\cite{Beenakker:1996ed, Beenakker:1996ch}, and obtained 21.2\,fb.  For $t\bar{t}+$jets, we used the NLO$+$NLL value calculated by the ATLAS and CMS Collaborations centrally using the {\tt Top++v2.0} package, which is 815.96\,fb for a top quark mass of 173.2\,GeV~\cite{Czakon:2011xx, Czakon:2013goa, Czakon:2012pz, Czakon:2012zr, Baernreuther:2012ws}.  Simulated events were weighted to correspond to 30\,fb$^{-1}$ of integrated luminosity.  The output {\tt ROOT} event files were analyzed using {\tt TNMAnalyzer}~\cite{tnm}.  

We then implemented the following object definitions.
\begin{itemize}
\item {\bf Jets: } anti-k$_T$ jets with jet size $R = 0.4$.  Jet $p_T > 30$\,GeV, $|\eta| < 2.4$.
\item {\bf b jets: } Jets defined as above, that pass the default {\tt Delphes} b jet tagger.  
\item {\bf W jets: } Cambridge-Aachen jets with jet size $R = 0.8$.  Jet mass of $60 < m_j < 120$\,GeV, n-subjettiness value~\cite{Thaler:2010tr} $\tau_2/\tau_1 < 0.5$.
\item {\bf Leptons ($\mathbf e$ and $\mathbf \mu$): } 
Lepton isolation $\sum_{\Delta R} p_T / p_T^{\ell} < 0.1$, where
$\Delta R < 0.5$, and lepton $p_T > 5$\,GeV, $|\eta| < 2.5$.
\item {\bf Missing transverse energy: } negative vector sum of transverse momenta of all visible objects.
\end{itemize}

\begin{figure}
\begin{center}
\includegraphics[width=0.32\textwidth]{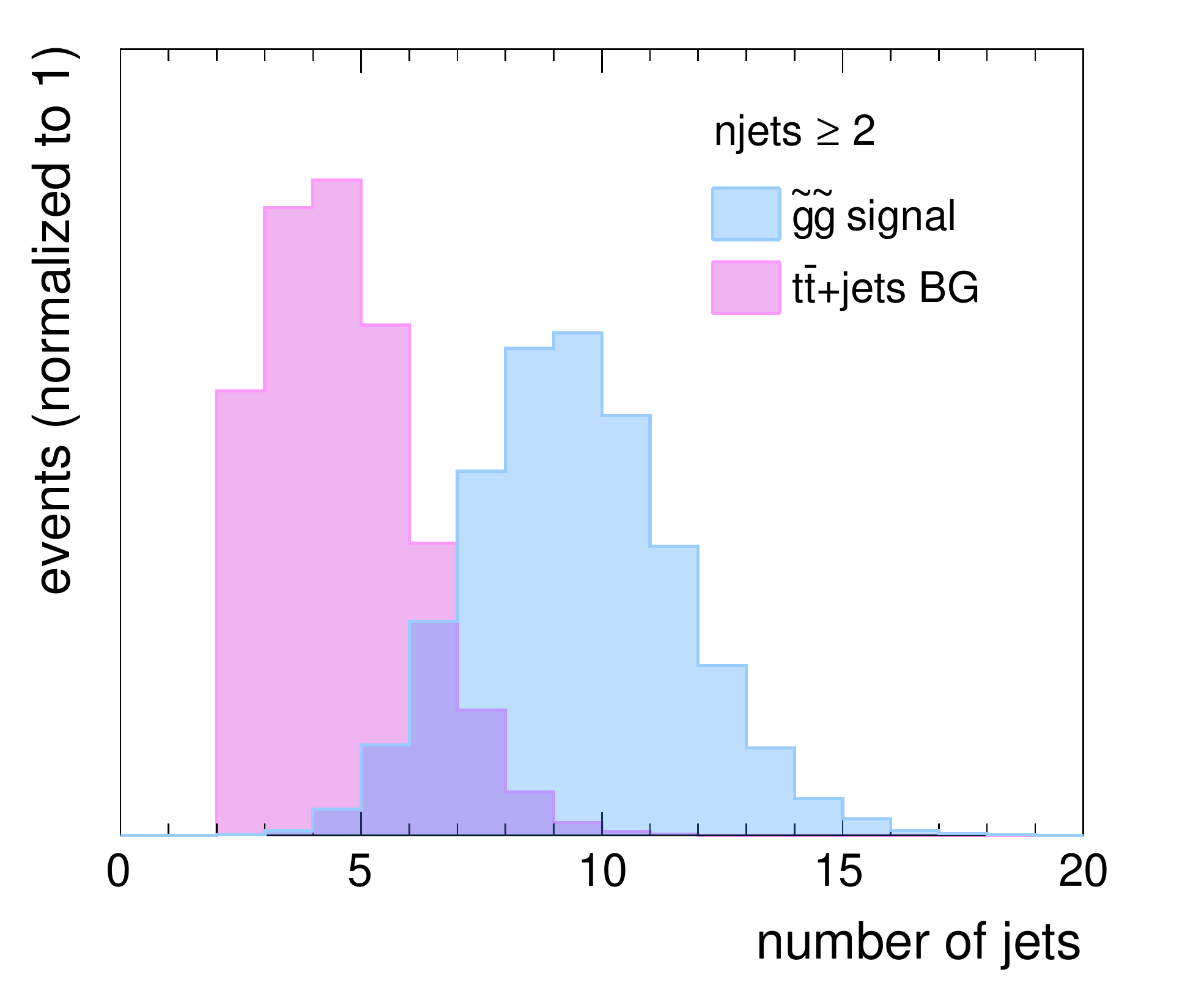}
\includegraphics[width=0.32\textwidth]{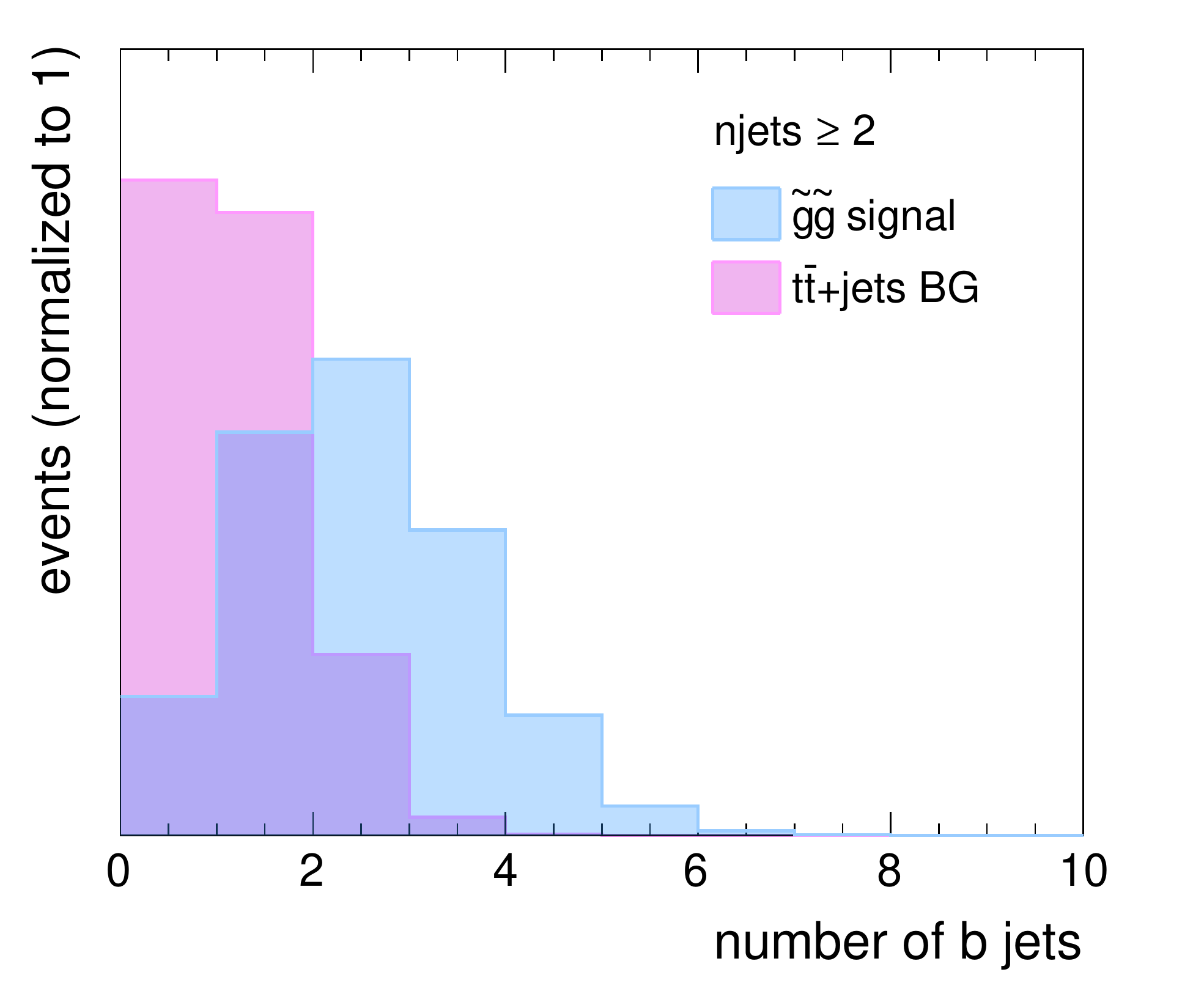}
\includegraphics[width=0.32\textwidth]{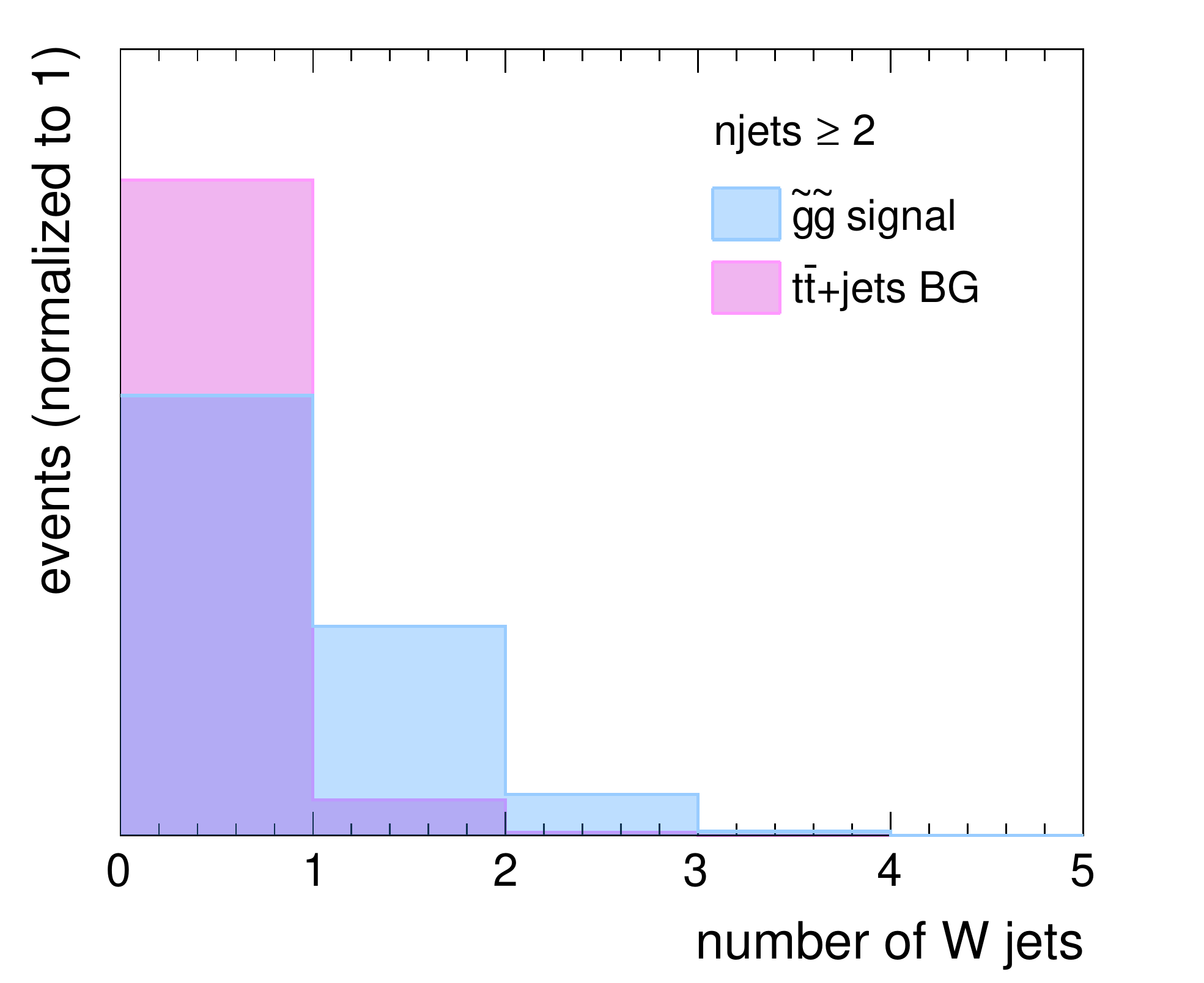} \\ 
\includegraphics[width=0.32\textwidth]{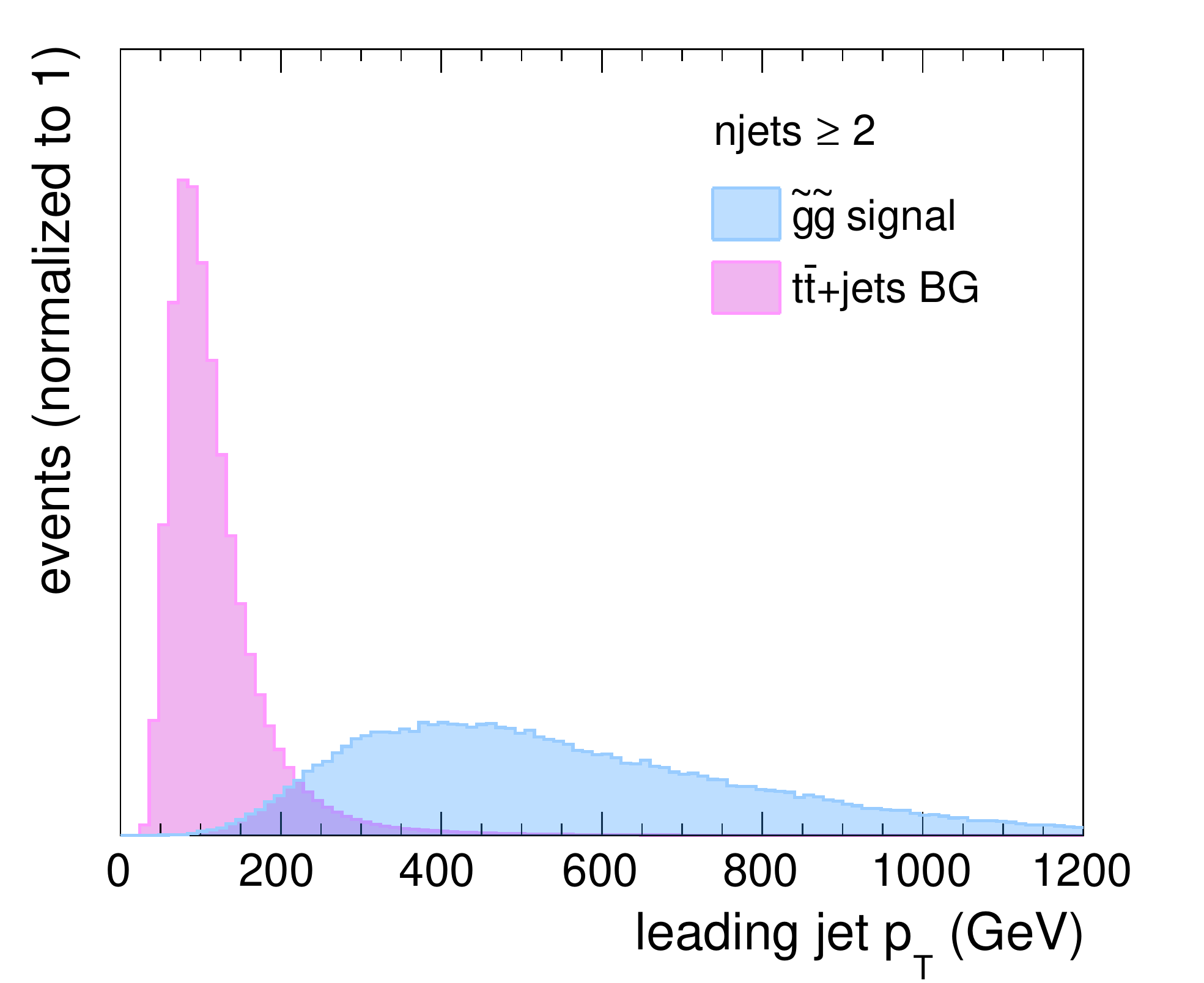}
\includegraphics[width=0.32\textwidth]{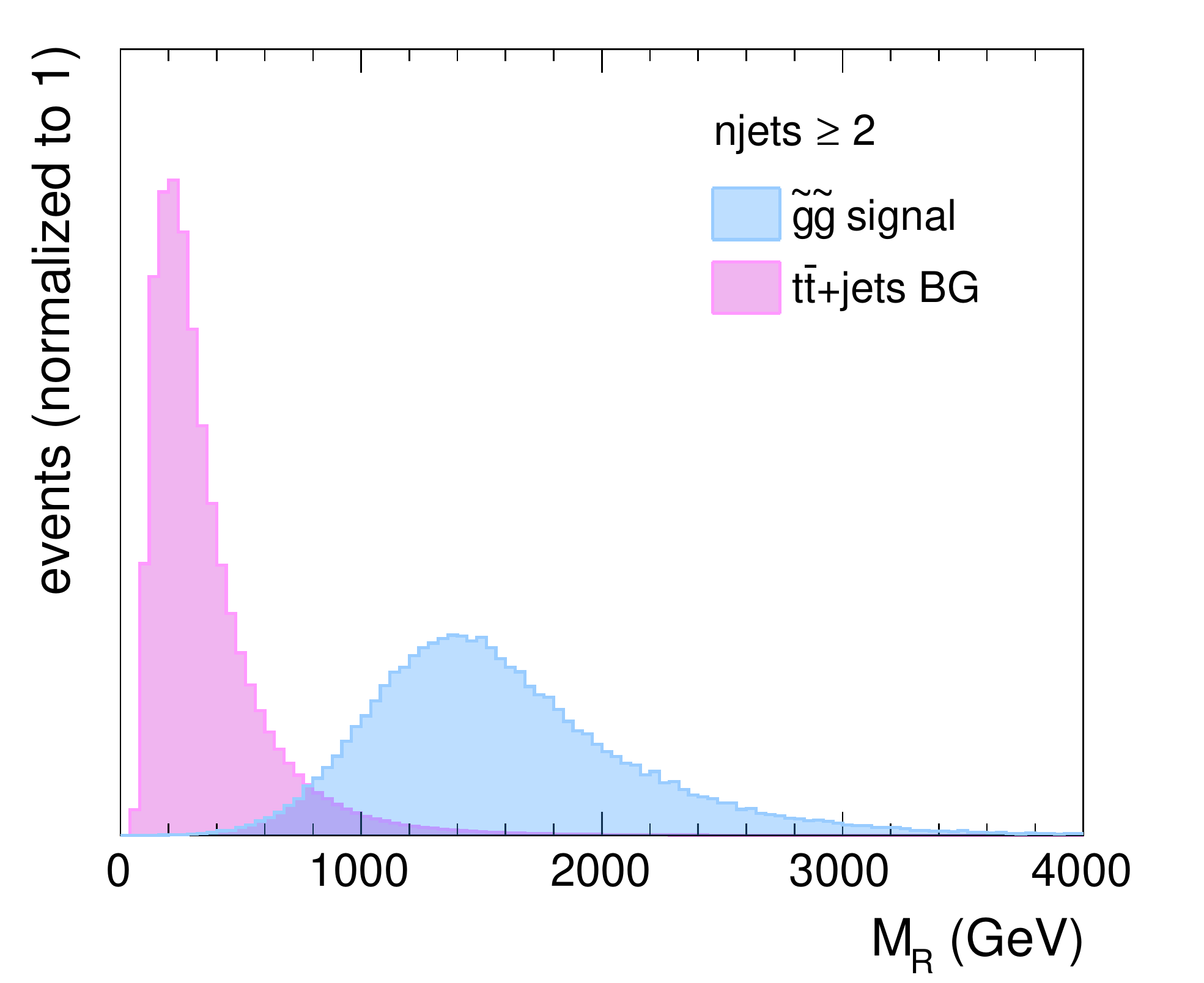}
\includegraphics[width=0.32\textwidth]{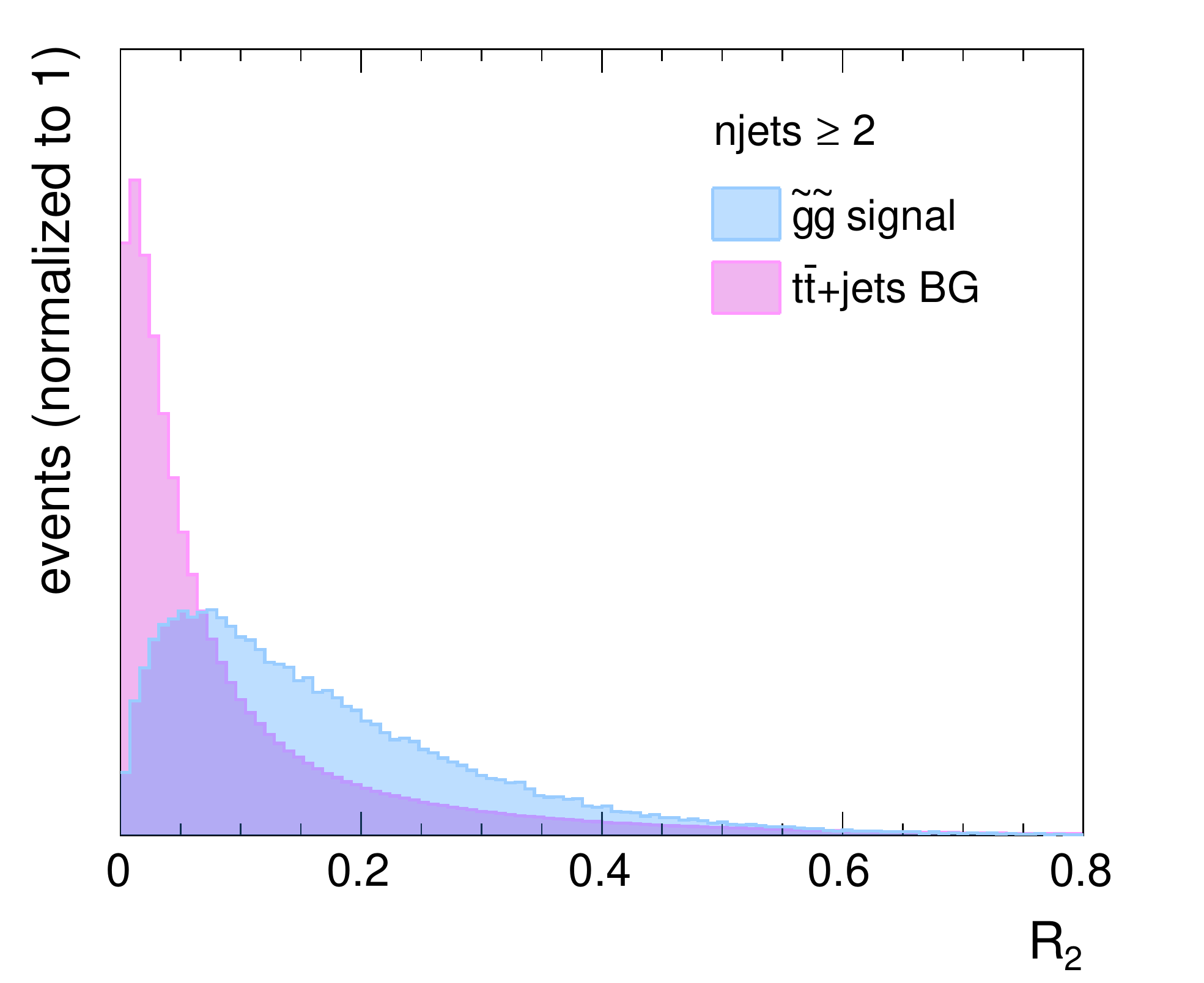} \\ 
\caption{Comparison of distributions for discriminating variables between the $\tilde{g}\tilde{g}$ signal and the $t\bar{t}+$jets background.  }
\label{fig:susyvarsshape}
\end{center}
\end{figure}

Based on these object definitions, events were selected that have at least 2 jets and no leptons.  For these events, we calculated the razor kinematic variables $M_R$ and $R^2$ as described in~\cite{Rogan:2010kb}.  For the RGS optimizations, we use the discriminating variable set consisting of the number of jets ($n_{j}$), number of b jets ($n_b$), number of W jets ($n_W$), leading jet $p_T$ ($p_T^{j1}$), $M_R$ and $R^2$.  Figure~\ref{fig:susyvarsshape} compares the distributions of these six variables for the SUSY signal and $t\bar{t}+$jets background, and confirms that each variable has indeed some measure of discrimination between the signal and the background.

The original CMS study selects events with $n_j \ge 3$, $n_b \ge 1$, $n_W \ge 1$ and $p_T^{j1} > 200$\,GeV (the ``preselection"), followed by the simple razor selection \mr$> 800$\,GeV and \r2$> 0.08$, which gives 140.1 signal events, 10702.4 background events, and a corresponding Z value of 1.35.  In the following, we optimize selections based on the above discriminant (cut) variables in order to improve the signal significance,
quantified by the Z value defined in Equation~(\ref{eq:Z}).  A summary of these optimization studies and their results is given in Table~\ref{tab:SUSopts}.  

\begin{figure}
\begin{center}
\includegraphics[width=0.49\textwidth]{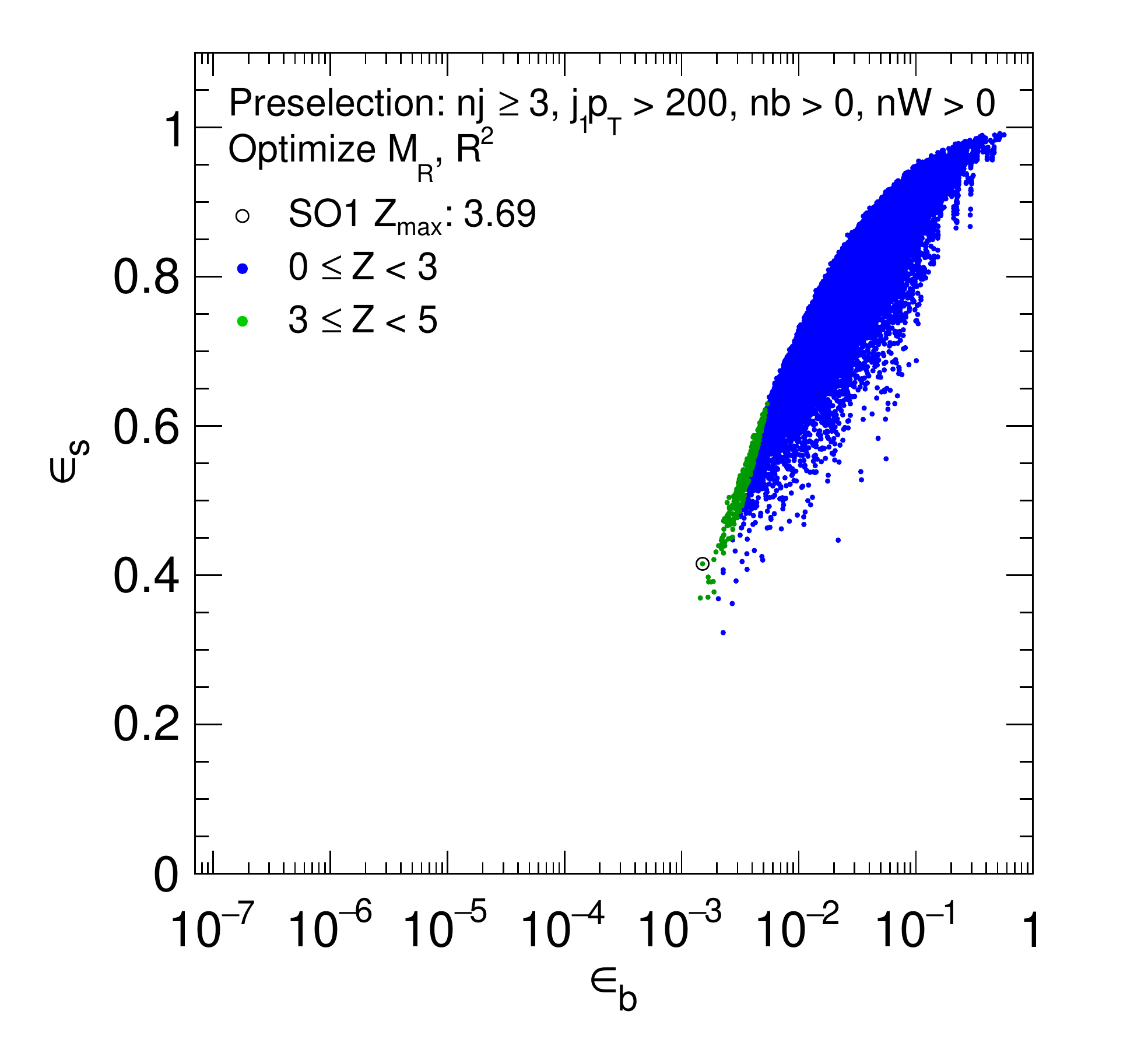}
\includegraphics[width=0.49\textwidth]{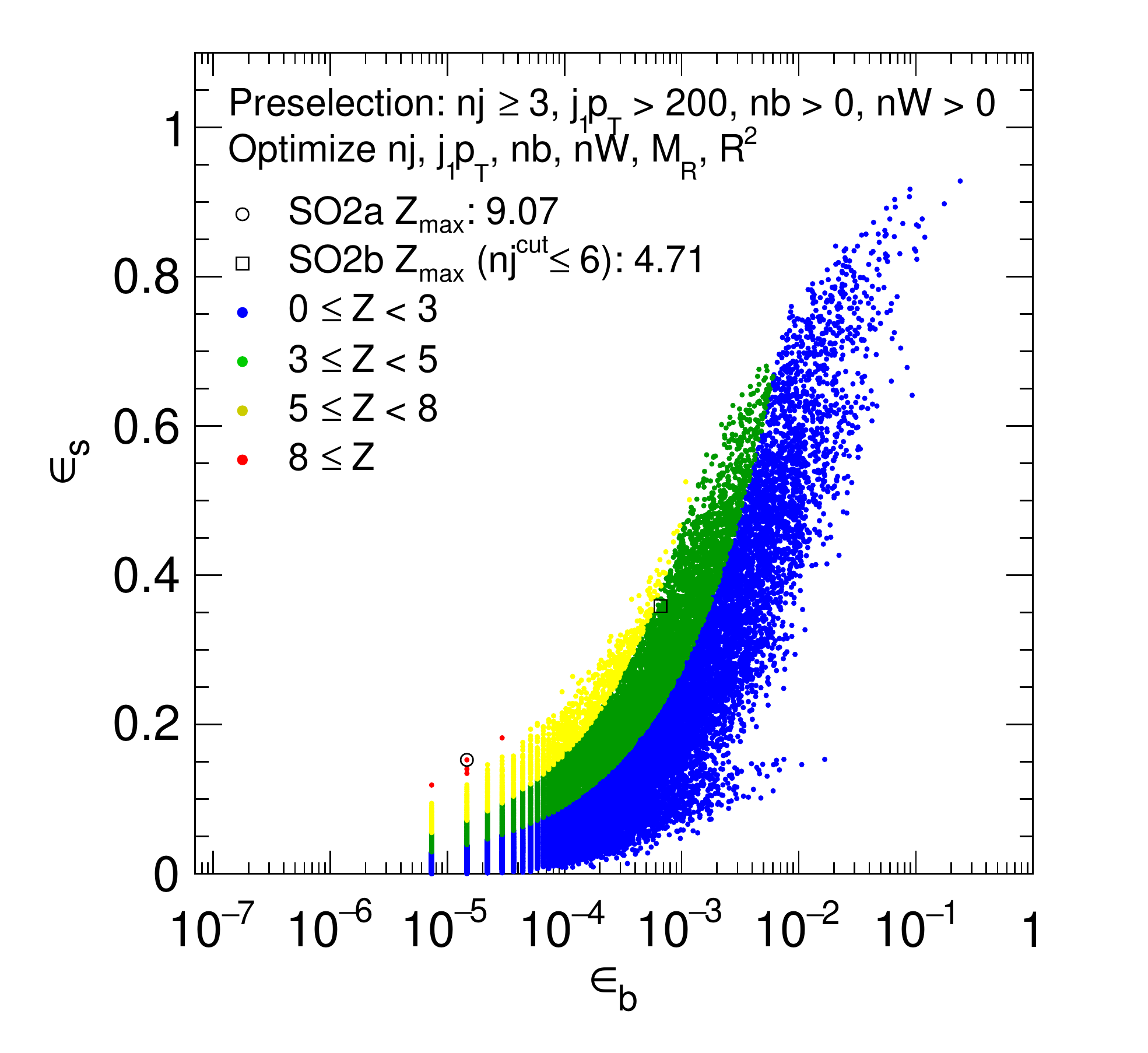} 
\includegraphics[width=0.49\textwidth]{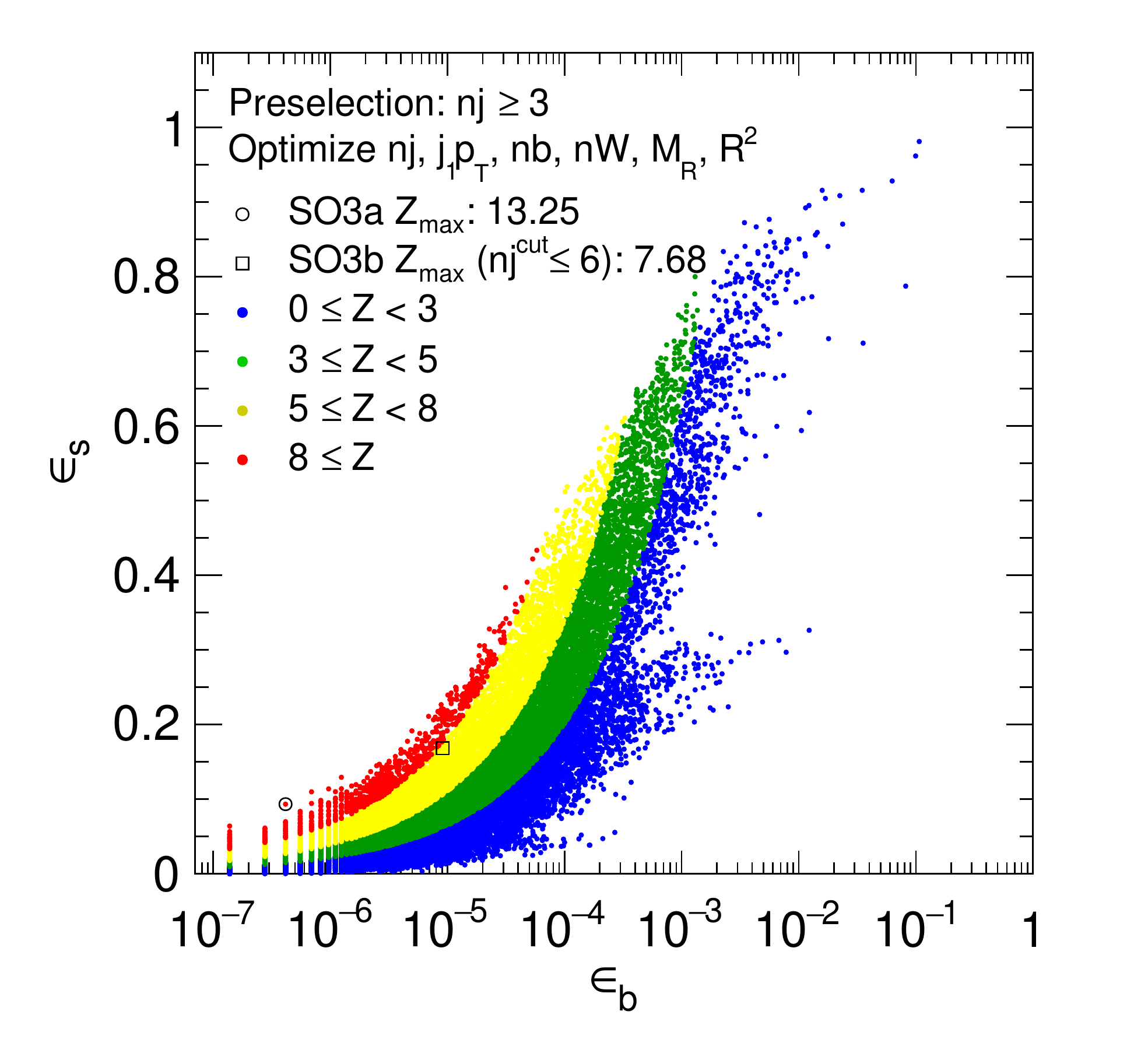} \\ 
\caption{Signal efficiency $\epsilon_s$ versus background efficiency $\epsilon_b$, and Z significance ranges for the three different classes of optimizations in this study: SO1 (top left), SO2a and SO2b (top right), SO3a and SO3b (bottom).  See further details in Table~\ref{tab:SUSopts}.}
\label{fig:SUSopts}
\end{center}
\end{figure}

\begin{figure}
\begin{center}
\includegraphics[width=0.49\textwidth]{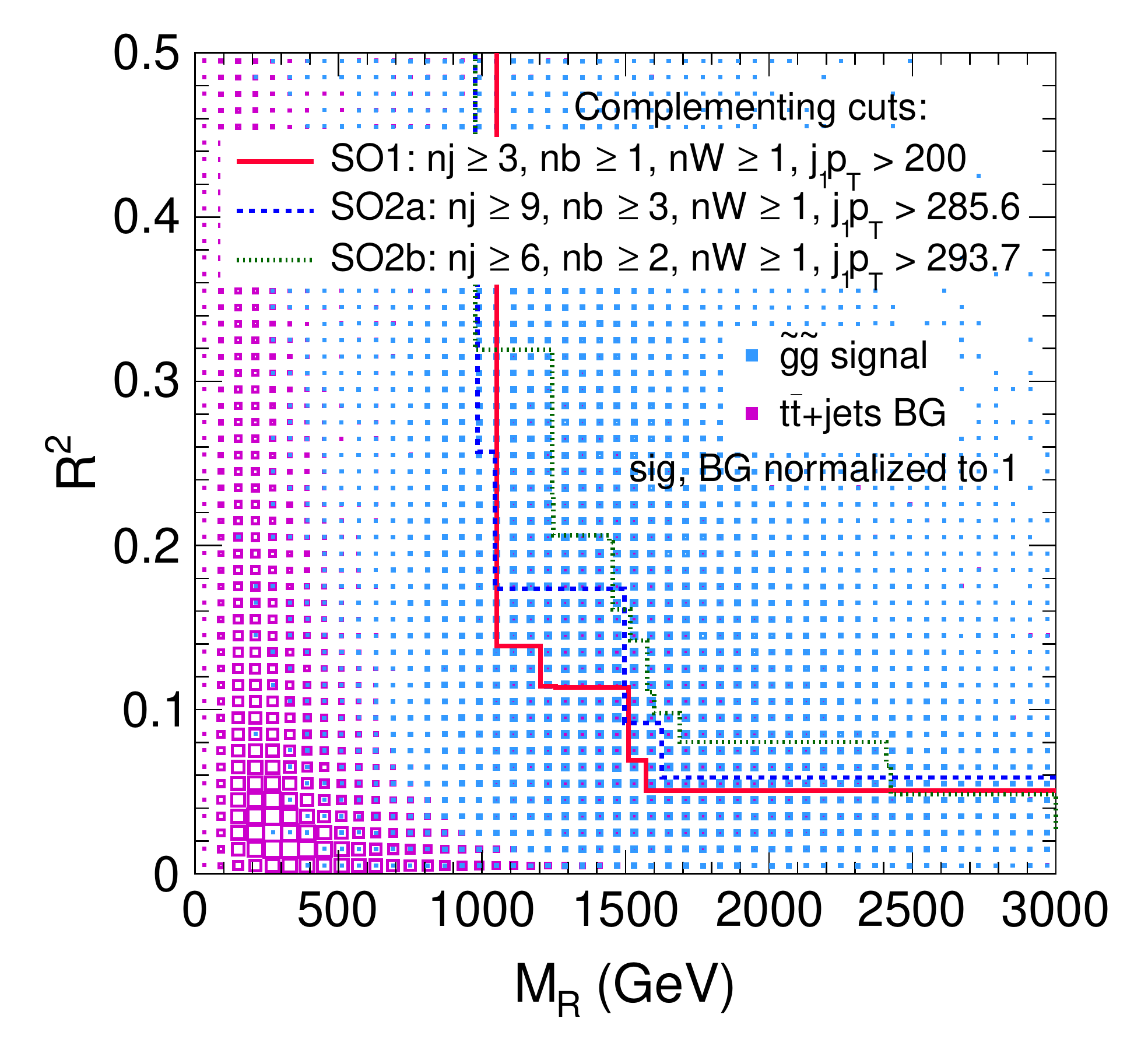}
\includegraphics[width=0.49\textwidth]{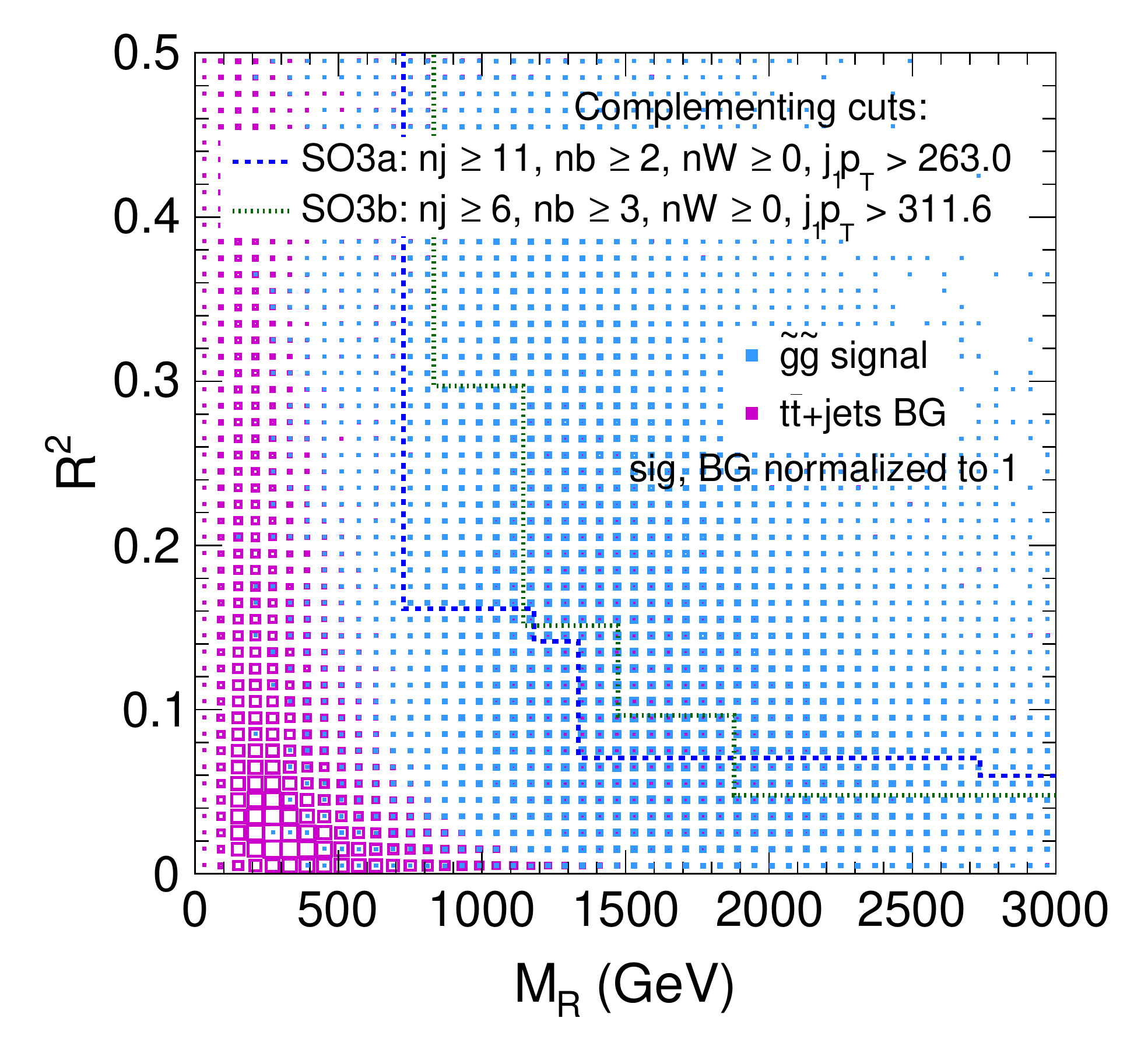}
\caption{\mr-\r2 cut boundaries determined by 20 step 2-dimensional staircase cuts in the 5 different optimizations in this study.  Cut boundaries are shown for optimizations with a preselection requiring (left) and not requiring (right) a boosted W boson.   
Both frames also show the \mr-\r2 distributions for the $\tilde{g}\tilde{g}$ signal and the $t\bar{t}+$jets background.}
\label{fig:MRR2opts}
\end{center}
\end{figure}

In the first SUSY optimization, SO1, we simply implement the CMS preselection and attempt to improve the razor selection through optimizing $M_R$ and $R^2$ using a 2-dimensional, 20 step, staircase cut.  The top left frame of Figure~\ref{fig:SUSopts} shows the signal efficiencies $\epsilon_s$ versus background efficiencies $\epsilon_b$ and Z significance intervals obtained for each candidate staircase cut.  As for the Higgs
studies, the cut with the largest Z significance is taken to be the optimal selection.  The optimal \mr-\r2 selection in this case has a Z value of 3.7, resulting from signal and background counts of $N_S = 85$ and $N_B = 504$ respectively. The associated cut is shown in the right frame of Figure~\ref{fig:MRR2opts} as the red curve.  

Next, in optimization SO2a, we again implement the preselection, but this time optimize over all six variables, with one-sided cuts on $n_j$, $n_b$, $n_W$, $p_T^{j1}$, and a 2-dimensional, 20 step staircase cut on \mr-\r2.  For this case, $\epsilon_s$ versus $\epsilon_b$ distributions and Z intervals are given in the top left frame of Figure~\ref{fig:SUSopts}, which shows that including all variables in the optimization can dramatically improve the selection performance.  The optimal selection with the best Z value of 9.07 is given in Table~\ref{tab:SUSopts} and the left frame of Figure~\ref{fig:MRR2opts}.  This selection includes a cut of $n_j \ge 9$, which restricts us to a high multiplicity phase space, where a fully reliable event reconstruction becomes harder to achieve.  Furthermore, this selection results in low event yields, which would lead to large uncertainties.  In order to overcome these issues, we look for a cut set where the cut value for $n_j$, denoted by $n_j^{cut}$, is at most 6.  The optimal selection in this case, SO2b, has $n_j\ge6$, a Z value of 4.72, and is shown in Table~\ref{tab:SUSopts} and the left frame of Figure~\ref{fig:MRR2opts}.  

\begin{table}
\caption{Summary of the different SUSY optimizations (SOs) and results for the $\tilde{g}\tilde{g}$ signal over the 
$t\bar{t}+$jets background for an integrated luminosity of $30\,\textrm{fb}^{-1}$.  $Z_{max}$ is the maximum value of the Z significance.
We use the significance measure defined in Eq.~(\ref{eq:Z}).}
\begin{center}
\begin{tabular}{|l|l|c|c|c|c|c|c|c|c|}
\hline
Opt & State & $n_j$ & $n_b$ & $n_W$ & $p_T^{j1}$ & \mr,\r2 & $N_S$ & $N_B$ & $Z_{max}$ \\
\hline
\hline
CMS & Analysis cuts ($var$) & $\ge 3$ & $\ge 1$ & $\ge 1$ & $> 200$ & \mr$>$800 & 140.1 & 10702.4 & 1.35 \\
         &  &  &  &  &  & \r2 $>$ 0.08 & & & \\
\hline
\hline
SO1 & Before opt. ($var$) & $\ge 3$ & $\ge 1$ & $\ge 1$ & $> 200$ & -- & \multicolumn{3}{|l|}{} \\
       & Optimized vars & $\times$ & $\times$ & $\times$ & $\times$ & $\checkmark$ & \multicolumn{3}{|l|}{} \\
       & After opt. ($var$) & $\ge 3$ & $\ge 1$ & $\ge 1$ & $> 200$ & Fig~\ref{fig:MRR2opts}, left & 85.2 & 504.2 & 3.69  \\
\hline
\hline
SO2a & Before opt. ($var$) & $\ge 3$ & $\ge 1$ & $\ge 1$ & $> 200$ & -- & \multicolumn{3}{|l|}{} \\
     & Optimized vars & $\checkmark$ & $\checkmark$ & $\checkmark$ & $\checkmark$ & $\checkmark$ & \multicolumn{3}{|l|}{} \\
     & After opt. ($var$) & $\ge 9$ & $\ge 3$ & $\ge 1$ & $> 285.6$ & Fig~\ref{fig:MRR2opts}, left & 31.3 & 4.9 & 9.07 \\
\hline
SO2b & Before opt. ($var$) & $\ge 3$ & $\ge 1$ & $\ge 1$ & $>200$ & -- & \multicolumn{3}{|l|}{} \\
     & Optimized vars & $\checkmark$ & $\checkmark$ & $\checkmark$ & $\checkmark$ & $\checkmark$ & \multicolumn{3}{|l|}{} \\
     & Cut value req. ($var^{cut}$) & 3-6 & -- & -- & -- & -- & \multicolumn{3}{|l|}{} \\
     & After opt. ($var$) & $\ge 6$ & $\ge 2$ & $\ge 1$ & $> 293.7$ & Fig~\ref{fig:MRR2opts}, left & 73.6 & 220.3 & 4.72 \\
\hline
\hline
SO3a & Before opt. ($var$) & $\ge 3$ & -- & -- & -- & -- & \multicolumn{3}{|l|}{} \\
     & Optimized vars & $\checkmark$ & $\checkmark$ & $\checkmark$ & $\checkmark$ & $\checkmark$ & \multicolumn{3}{|l|}{} \\
     & After opt. ($var$) & $\ge 11$ & $\ge 2$ & $\ge 0$ & $> 263$ & Fig~\ref{fig:MRR2opts}, right & 59.4 & 7.3 & 13.25 \\
\hline
SO3b & Before opt. ($var$) & $\ge 3$ & -- & -- & -- & -- & \multicolumn{3}{|l|}{} \\
     & Optimized vars & $\checkmark$ & $\checkmark$ & $\checkmark$ & $\checkmark$ & $\checkmark$ & \multicolumn{3}{|l|}{} \\
     & Cut value req. ($var^{cut}$) & 3-6 & -- & -- & -- & -- & \multicolumn{3}{|l|}{} \\
     & After opt. ($var$) & $\ge 6$ & $\ge 3$ & $\ge 0$ & $> 312$ & Fig~\ref{fig:MRR2opts}, right & 107.1 & 161.6 & 7.68 \\
\hline
\hline
\end{tabular}
\end{center}
\label{tab:SUSopts}
\end{table}%

The above study with the requirement of at least one boosted $W$ boson already demonstrates the power of the RGS algorithm in finding diverse options in improving event selections.  
The RGS algorithm can also be used for easy exploration of alternative signatures for a given signal.  To illustrate this, we start with a much simpler preselection of $n_j \ge 3$, and optimize over all six cut variables.  Figure~\ref{fig:SUSopts} (lower frame) shows that the more open phase space makes it possible to reach much lower $\epsilon_b$ for a given $\epsilon_s$, and consequently higher Z significances.  The optimal selection for the most generic optimization, denoted as SO3a, is given in Table~\ref{tab:SUSopts} and Figure~\ref{fig:MRR2opts}, and has a very high Z significance of 13.25.  This optimization is achieved when the requirement $n_W \ge 1$ is lifted, but it has a very high jet multiplicity requirement of $n_j \ge 11$, which brings with it the same issues as in the SO2a optimization.  
To avoid these issues, we again choose the best cut set for which $n_j^{cut} \le 6$, SO3b, and which has a Z significance of 7.68 and cut values shown in Table~\ref{tab:SUSopts} and Figure~\ref{fig:MRR2opts} right frame.  

\begin{figure}
\begin{center}
\includegraphics[width=0.32\textwidth]{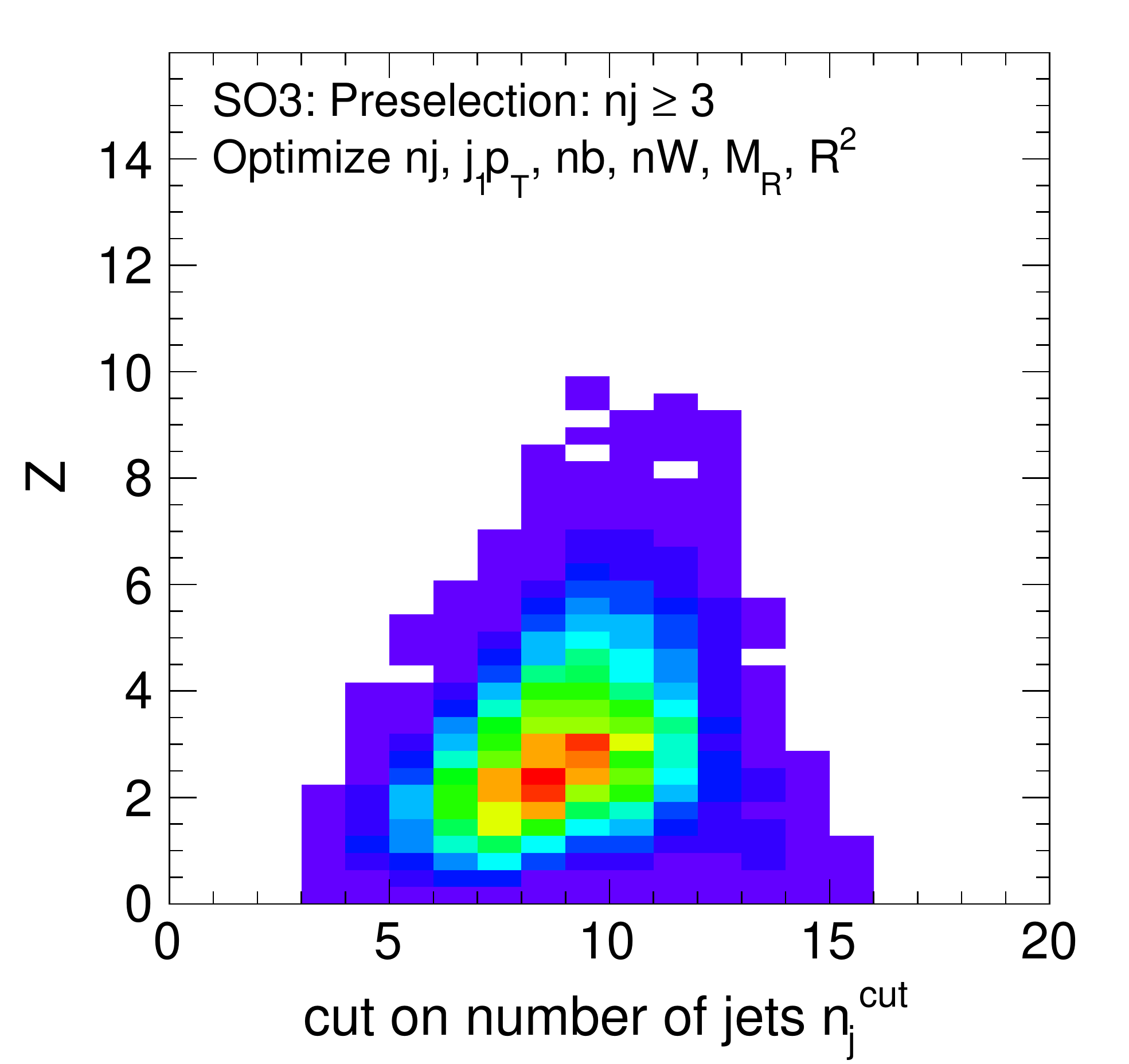}
\includegraphics[width=0.32\textwidth]{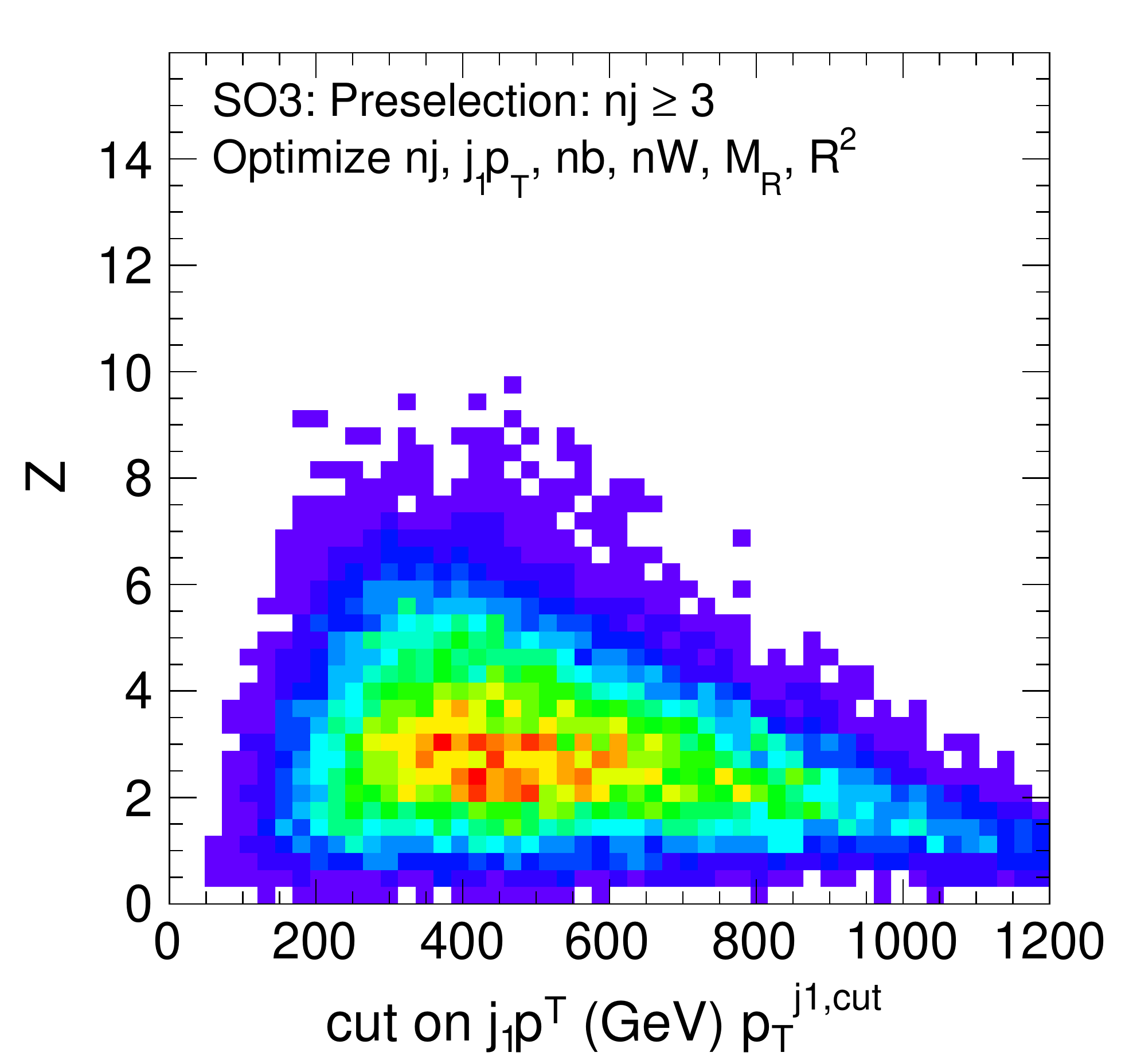} \\
\includegraphics[width=0.32\textwidth]{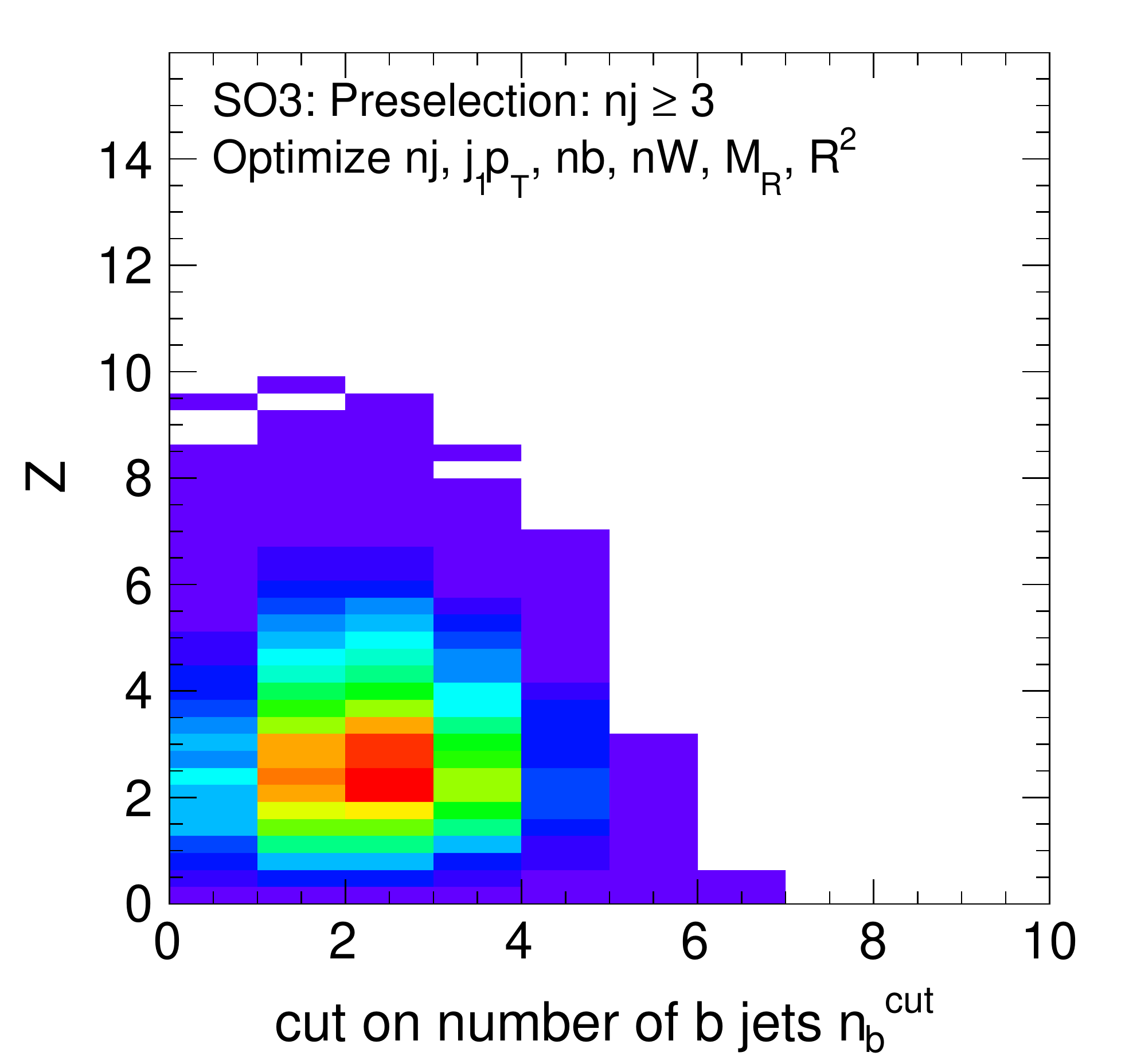}  
\includegraphics[width=0.32\textwidth]{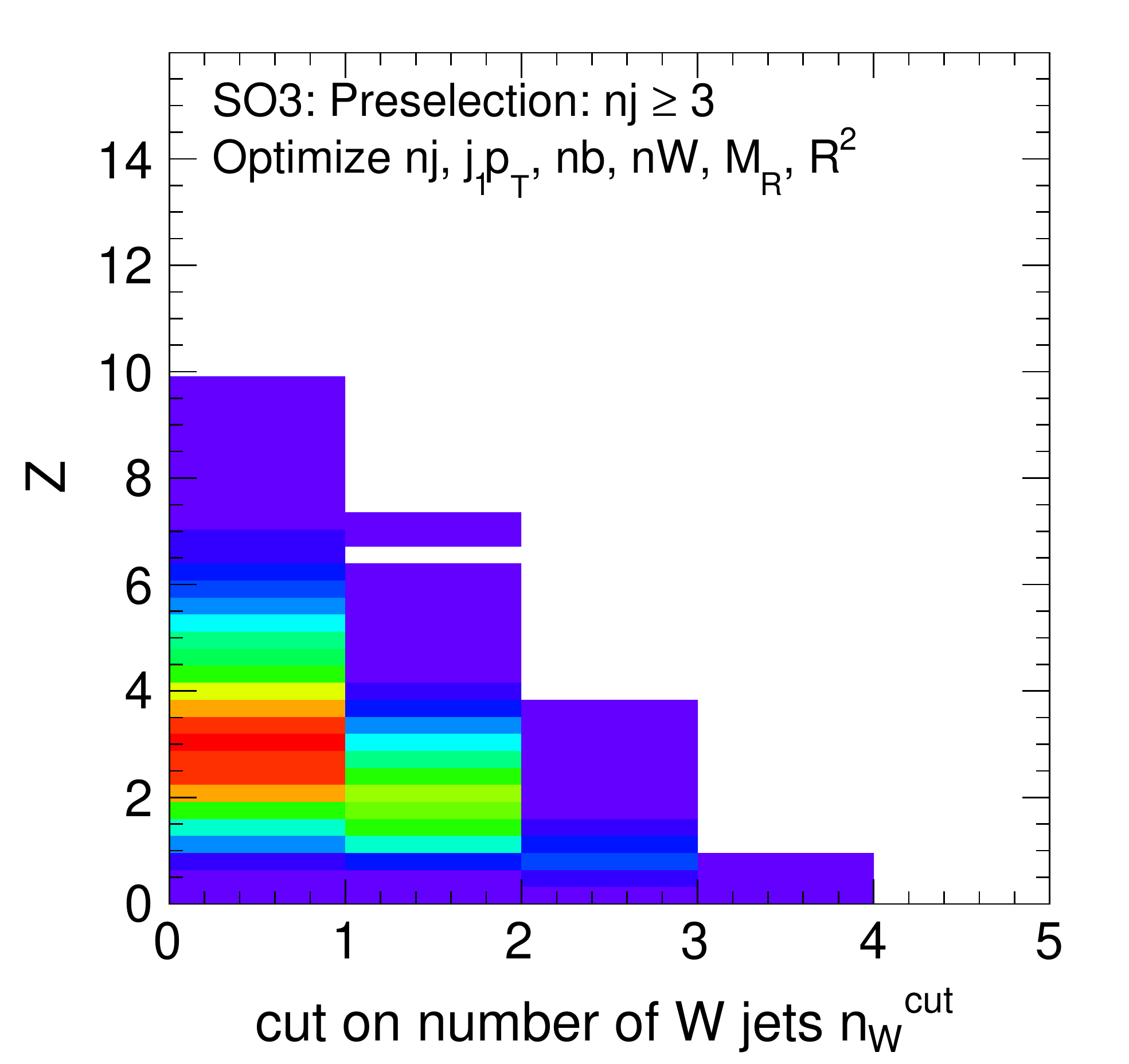}  
\caption{2-dimensional histograms of number of cut sets for Z significance versus $n_j^{cut}$ (top left), $p_T^{j1, cut}$ (top right), $n_b^{cut}$ (bottom left) and $n_W^{cut}$ (bottom right) for the SO3-type optimization. }
\label{fig:SUSYvarZ}
\end{center}
\end{figure}

\begin{figure}
\begin{center}
\includegraphics[width=0.32\textwidth]{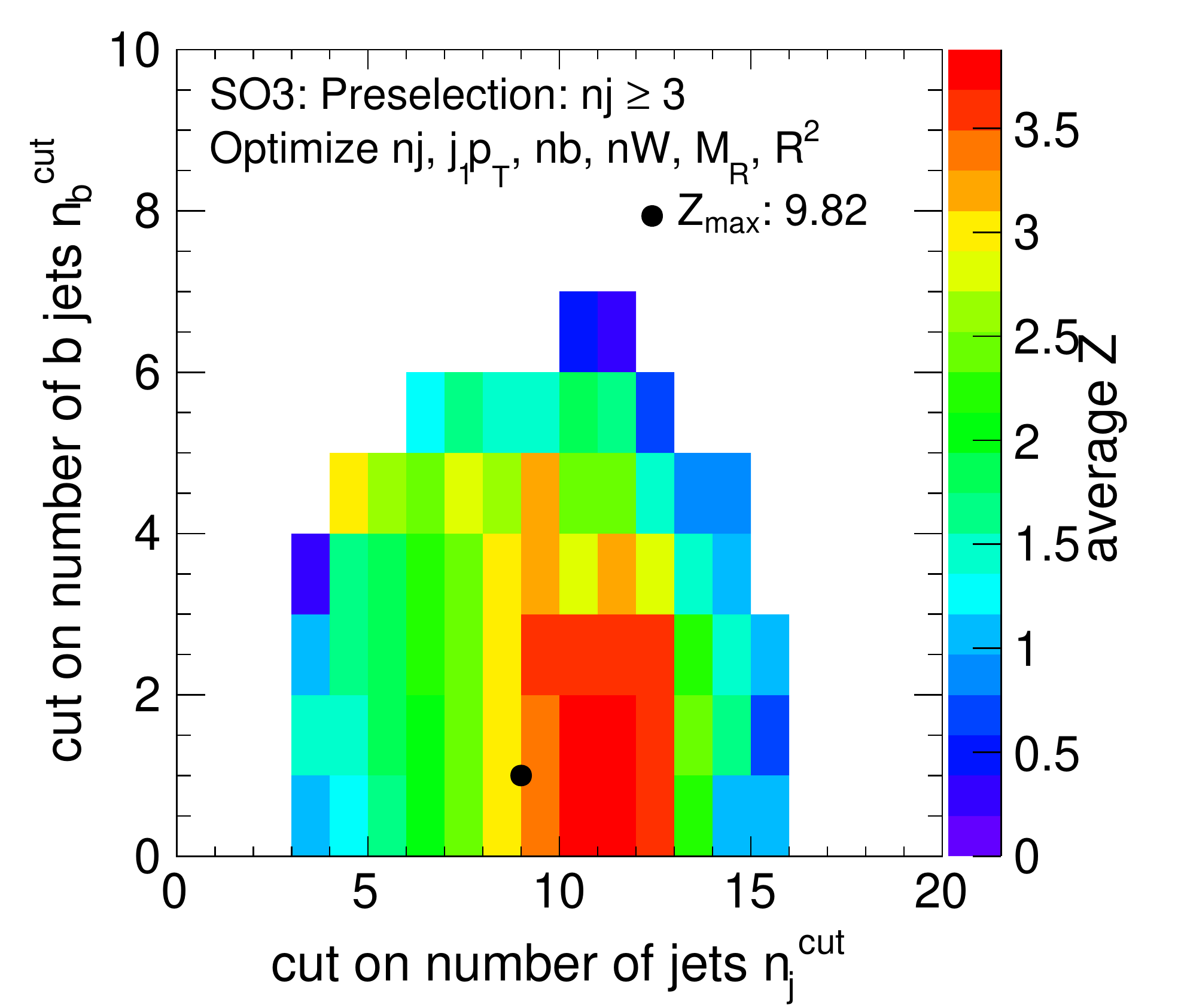}
\includegraphics[width=0.32\textwidth]{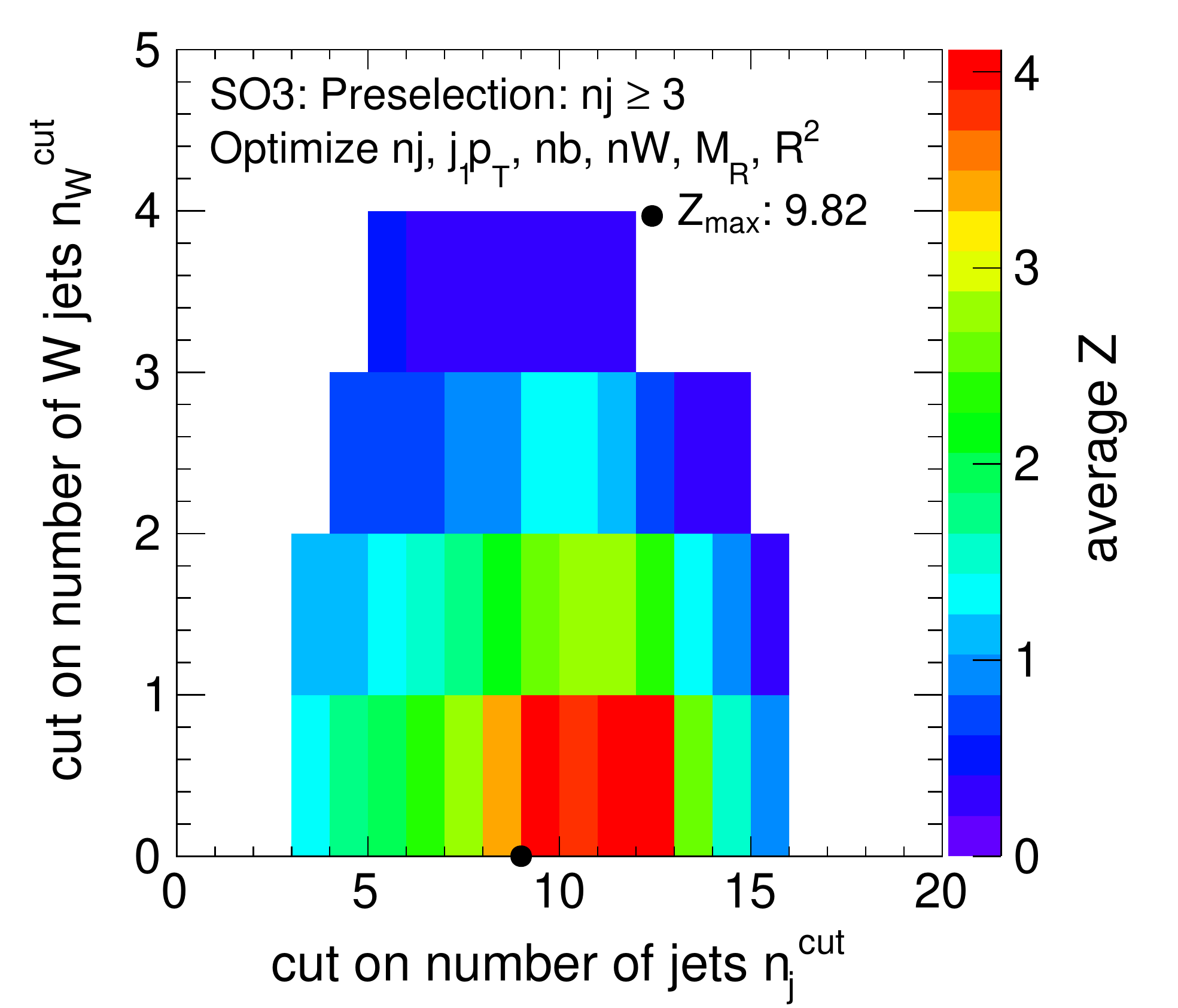} 
\includegraphics[width=0.32\textwidth]{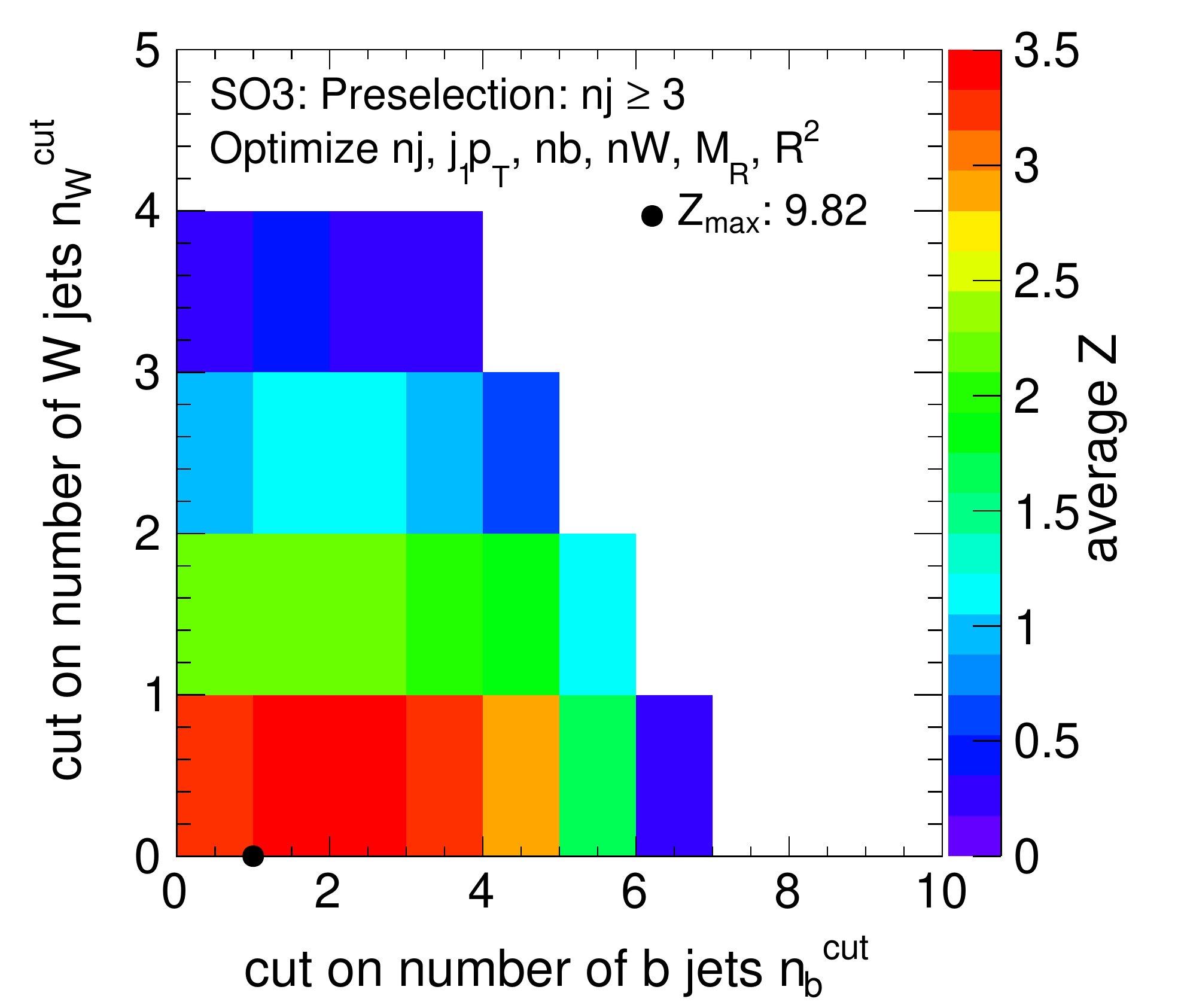} \\
\caption{The average value on the $z-$axis versus cut values of different cut variables on $x-$ and $y$ axes.}
\label{fig:SUSYvarsavZ}
\end{center}
\end{figure}

The RGS optimization can be very helpful in finding complementary approaches when searching for a given signal.  Here, we have shown that the $\tilde{g}\tilde{g} \rightarrow t\tilde{t}_1t\tilde{t}_1 \rightarrow tt\tilde{\chi}_1^0tt\tilde{\chi}_1^0$ signal can be explored with the razor kinematic variables in final states with boosted W bosons, or in generic hadronic final states with $b$ jets.  Overall, RGS showed that 
final states with high jet multiplicities (detector conditions permitting) are the most sensitive to new physics,
independently of whether or not  one requires the presence of boosted W bosons.  Comparing the left and right frames of Figure~\ref{fig:MRR2opts} shows that selections with boosted $W$ bosons favor higher $M_R$ values.  RGS can be easily used for learning the characteristics of the cut phase space.  For example, the four frames in Figure~\ref{fig:SUSYvarZ} show the 2-dimensional histograms of the number of cut sets for Z significance versus $n_j^{cut}$, $p_T^{j1, cut}$, $n_b^{cut}$ and $n_W^{cut}$ respectively for SO3.  These histograms show that the highest Z values can be obtained for $n_j^{cut} \sim 10$, $p_T^{j1, cut} \sim$ 200-300, $n_b^{cut} \sim 2-3$, and $n_W^{cut} = 0$.  Similarly, Figure~\ref{fig:SUSYvarsavZ} shows the average Z significance value on the $z-$axis versus cut values of different cut variables on the $x-$ and $y$ axes, which could help to characterize the best cut value combinations for different variables.

\section{Summary}
\label{sec:summary}

We have described a simple algorithm called the random grid search that  efficiently optimizes event selections by randomly searching for cuts in the region of the cuts phase space where the cuts are  likely to be most useful.  Furthermore, we have described recent enhancements to RGS, specifically, the addition of two-sided cuts, and a cut type we call a staircase cut that is the OR of two or more  sets of one-sided cuts on two or more variables.  Next, we presented two examples which showed the effectiveness of RGS optimization in exploring the impact of different cuts on different sets of variables.

It is undoubtedly true that for the most difficult analyses, in which optimality is at a premium, multivariate
discriminants constructed using machine learning will likely win the day over any cut-based analysis. 
However, that expectation does not obviate the utility of optimal cut-based analyses.  
A significant fraction of LHC analyses are still cut-based.  Feasibility studies of almost all searches are done using cut-based analyses.  
Furthermore, there are circumstances in which a region enhanced in some way is needed, perhaps a control region for an
analysis, which is to be determined by cuts applied to one or more variables.
These variables could themselves be sophisticated multivariate discriminants. Then,
a simple, effective, way to find the optimal cuts, such as the random grid search, would be useful.  There are, of course,
other algorithms to search for optimal cuts. One that has found some favor in particle physics finds
cuts using a genetic algorithm. However, we would argue that sophistication of this kind is unnecessary for
most cut-based analyses and an algorithm like RGS is perfectly adequate.  Moreover, it is easy to
envisage further useful developments. Indeed, one obvious development would be to extend the staircase cut to permit the use of two-sided cuts that will make it possible to find the OR of quite general multi-dimensional 
bounded regions. Work towards this is in progress.

\section*{Acknowledgements}

The work of PCB and HBP is supported in part by the U.S. Department of Energy, under contract number DE-AC02-07CH11359 with Fermilab, and under grant number DE-SC0010102, respectively.  The work of SS is supported by the financial support of the National Research Foundation of Korea (NRF), funded by the Ministry of Science \& ICT under contract NRF-2008-00460 and by the U.S. Department of Energy through the Distinguished Researcher Program from the Fermilab LHC Physics Center.

\appendix
\section{Using RGS}
\label{sec:manual}


The RGS algorithm comprises two independent steps. The first imposes a set of event selections based on selection variables and cut directions defined by the user, and corresponding cut values taken from a set of specified events (usually signal events), and it determines the resulting signal and background efficiencies and counts for each selection.  These efficiencies and counts are then used in the 
second step to compute a measure of signal significance, which can be maximized over different selections in order to determine the optimal event selection.  

The efficiency of the algorithm derives from the fact that the search for cuts is stochastic and
determined by (in most applications) the predicted signal distribution. In effect, this distribution provides an importance sampled set of cuts. Consequently, the algorithm is less subject to the
``curse of dimensionality" and the accuracy of the results, as is true of Monte Carlo
integration, is independent of the dimensionality of the cut space.  The accuracy rather depends on
the sample sizes. The examples released with the package run in minutes. But, since the time
to run is roughly proportional to the product of the number of cuts and the number of events to
which the cuts are applied, for large sample sizes the RGS algorithm can take  several hours. However,
RGS can be trivially parallelized simply by splitting the events to which cuts are to be applied into
many batches, running RGS in parallel, and, for each cut, summing the signals and backgrounds from
each batch.

Running the RGS package requires {\tt C++}, {\tt Python} and {\tt ROOT} with {\tt PyROOT} built in.  The package can be obtained from {\tt github} and compiled as follows.
\begin{code}
git clone https://github.com/hbprosper/RGS.git
cd RGS
source setup.(c)sh
make
\end{code}
The package is released with code for the two examples described in this paper, in 
the directory tree
\begin{verbatim}
RGS/examples
            /data
            /Higgs
                  /HO1
                  /HO2
                  /HO3
            /SUSY
                  /SO1
                  /SO2
                  /SO3
\end{verbatim}
The {\tt ROOT} files used in the Higgs and SUSY examples should be downloaded and unpacked into the {\tt data} directory using, for example, the commands
\begin{code}
wget http://www.hep.fsu.edu/~harry/RGS/data/SUSY.tar.gz
tar zxvf SUSY.tar.gz
\end{code}

The package contains a single {\tt C++} class called {\tt RGS} that is callable from {\tt Python}.  
Instructions for running the examples are provided in the {\tt README} file in the {\tt examples}
directory. For each example, the event variables and cut directions to be used for optimization are specified in a text file with a simple syntax.  The signal and background events can be input in simple text format or as {\tt ROOT} trees. Using {\tt ROOT} trees has the significant advantage that it is possible to apply selection criteria to the events that are to be used in the search for cuts.

%

Below we give a more detailed description of the variables and cut directions file and the RGS class.

\subsection{Variables and cut directions file}

The variables and cut directions to be used in the optimization are input to RGS through a text file with a specific syntax.  The variable names should correspond to the variable names in the cut and search datafiles.  Currently RGS allows optimization with three types of cuts, which could be input as follows:
\begin{itemize}
\item {\bf One-sided cuts:} These are simple cuts like $p_T > x$, $|\eta| \le y$, etc.  The syntax is:
\begin{code}
variable-name   cut-direction
\end{code}
The available types are:
\begin{itemize}
\item {\tt var < } : $var < value$
\item {\tt var > } : $var > value$
\item {\tt var <= } : $var \le value$
\item {\tt var >= } : $var \ge value$
\item {\tt var <| } : $|var| < value$
\item {\tt var >| } : $|var| > value$ 
\item {\tt var == } : $|var| = value$ 
\end{itemize}
\item {\bf Two-sided cuts:} These are cuts with two boundaries, as in the case of $x < m_{\ell\ell} < y$.  The two boundaries are taken from two separate events in the signal event list.  The syntax is
\begin{code}
variable-name  cut-direction
\end{code}
The two-sided cut type syntax is:
\begin{itemize}
\item {\tt var <>} : $value1 < var < value2$
\end{itemize}
\item {\bf Staircase cuts:} (initially inspired by the razor kinematic variables $M_R$ and $R^2$): In 2-dimensional space, this cut can be visulized as a sideview of a non-uniform staircase.  The staircase cut is constructed by selecting $N$ events from the signal, and for each event (or cutset $i$), taking the AND of a set of $M$ one-sided cuts.  The selection becomes the OR of the phase space covered by the $N$ cutsets.  The syntax is as follows:
\begin{code}
{\textbackslash}staircase number-of-steps N (i.e., cut-points)
    variable-name-a cut-direction-a
    variable-name-b cut-direction-b
    ...
    variable-name-M cut-direction-M
{\textbackslash}end
\end{code}
and a concrete example can be written as
\begin{code}
{\textbackslash}staircase N
        var_a  >
        var_b  >
        ...
        var_M  <
{\textbackslash}end
\end{code}
which would give

$(var_a > value_a^1 \,\, and \,\, var_b > value_b^1 \dots \, and \,\, var_M < value_M^1) \,\, or$ \\
$(var_a > value_a^2 \,\, and \,\, var_b > value_b^2 \dots \, and \,\, var_M < value_M^2) \,\, or$ \\
...  \\
$(var_a > value_a^N \,\, and \,\, var_b > value_b^N \dots \, and \,\, var_M < value_M^N)$. 

The staircase cut construction can accommodate cut types $<$, $>$, $<=$, $>=$, $<|$ and $|>$.

\end{itemize}

Below are some example variable and cut direction inputs used in this study:

\begin{code}
# Simple one-sided cuts
njet >=
j1pT >

# Two-sided cut
Z2mass <>

# Staircase cut
{\textbackslash}staircase 4
        MR  >
        R2  >
{\textbackslash}end
\end{code}
The cut definition files used in the Higgs and SUSY examples in this paper are named
\begin{code}
examples/Higgs/HOi.cut, i = 1, 2, 3
 \end{code}
and
\begin{code}
examples/SUSY/SOi.cut, i = 1, 2, 3
 \end{code}

\subsection{The RGS class}
\label{sec:rgsclass}

The RGS class is found in {\tt src/RGS.cc} and {\tt include/RGS.h}.  The functionalities and methods of the class are as follows.

\begin{itemize}
\item Constuctors:
\begin{code}
  RGS(std::string (or std::vector<std::string>&) cutdatafilename, 
      int start=0, 
      int numrows=0,
      std::string treename="",
      std::string weightname="",
      std::string selection="");
\end{code} 
where
\begin{itemize}
\item {\tt cutdatafilenames} are the names of one or more files containing the cut values, 
usually files with signal events.  The files can be in {\tt ROOT} or text format as described above.  
In a {\tt Python} program, multiple file names can be specified in a vector of strings using the {\tt PyROOT} wrapper {\tt vector(`string')}.
\item {\tt start} is the row or event where we start reading the cut values.
\item {\tt numrows} is the number of rows to be selected from the cuts file;  $-1$ specifies
that all rows are to be read.  The number of rows read will in general  be greater than the number of
selected rows if a selection has been applied when using a {\tt ROO}T n-tuple.
\item {\tt treename} is the name of the {\tt ROOT} tree to be read.  Tree names should be the same for the signal and background events used in RGS.  If omitted, the file to be read is presumed to be a text file.
\item {\tt weightname} is an optional variable that gives the name of the {\tt ROOT} branch or text column that contains the event weight.
\item {\tt selection} is an optional selection string to be used for selecting rows.  This is a {\tt ROOT} 
feature and so can only be used with {\tt ROOT} files.
\end{itemize}

\item Add a search file, or a list of searchfiles, which contain the signal and background event data:
\begin{code}
  void  add(std::string (or std::vector<std::string>&) searchfilename,
            int start=0,
            int numrows=-1,
            std::string resultname="",
            double weight=1.0);
\end{code}
where
\begin{itemize}
\item {\tt searchfilename} are the names of one or more files with signal and background data.  
Again, multiple files can be specified in {\tt Python} programs using the {\tt PyROOT} {\tt vector} 
wrapper.  All files should be in the same format.
\item {\tt start}, {\tt numrows} as described above.
\item {\tt resultname} is appended  to the count and fraction variables.  e.g. ,``\_s" for signal, ``\_b" for background.  If omitted, the ordinal value of the search file, starting at zero, is appended to
the count and fraction variables.
\item {\tt weight} is an optional (multiplicative) event weight assigned per search file.
\end{itemize}
\item Run the RGS algorithm for specified cut variables and cut directions: 
This can be done either using a cut definition file, which was described above:
\begin{code}
  void  run(std::string  varfile,  // file name of Variables file
            int nprint=500);
\end{code}
or by directly specifying cuts as strings:
\begin{code}
  void  run(vstring&  cutvar,  // Variables defining cuts
            vstring&  cutdir,  // Cut direction (cut-type)
            int nprint=500);
\end{code}
\item Return results:
\begin{itemize}
\item Return the total (possibly weighted) count for the data file identified by {\tt dataindex}:
\begin{code}
  double    total(int dataindex);
\end{code}
\item Return the (possibly weighted) count for the given data file and the given cut-point:
\begin{code}
  double    count(int dataindex, int cutindex);
\end{code}
\item Return all variables read from the cut file(s):
\begin{code}
  vstring&  vars();
\end{code}
\item Return number of cuts:
\begin{code}
  int       ncuts();
\end{code}
\item Return cut values for cut-point identified by cutindex.
\begin{code}
  vdouble   cuts(int cutindex);
\end{code}
\item Return cut variable names
\begin{code}
  vstring   cutvars();
\end{code}
\item Return number of events for a specified data file.
\begin{code}
  int       ndata(int dataindex);
\end{code}
\item Return values for data given data file and event.
\begin{code}
  vdouble&  data(int dataindex, int event);
\end{code}
\end{itemize}
\item Save resulting counts and fractions to a text or a {\tt ROOT} file with name {\tt filename} (if {\tt .root} is found in file name, a {\tt ROOT} file is saved, otherwise, a text file is saved):
\begin{code}
  void    save(std::string filename);
\end{code}  
\end{itemize}

\subsection{Running RGS}
The examples released with the RGS package show how to use the RGS class and how to analyze its results.  The examples use {\tt Python}, but
the {\tt RGS} class can be used in either a {\tt C++} or {\tt Python} program.  

Each example can be run by executing the command 
\begin{code}
./train.py
\end{code}
which instantiates an {\tt RGS} object and calls the relevant methods to add signal and background files, run the {\tt RGS} algorithm, and write the results to a {\tt ROOT} file. (Note that due to the length of the SUSY signal and background datasets, the SO3 optimization could take more than an hour. 
Here is a clear case in which splitting the search files into multiple files and running RGS in 
parallel would speed things up considerably.
SO1
and SO2 each take about 10 minutes.)
The results  in the {\tt ROOT} file can be analyzed by running 
\begin{code}
./analysis.py
\end{code}
which calculates the $Z$ value defined in Eq.~(\ref{eq:Z})  for each cut and finds the
best cut (the one with the largest $Z$ value).  A ROC curve of signal and background efficiencies is also
plotted.  
A collection of helpful utility functions may be found in {\tt python/rgsutil.py}, which can be imported using
\begin{code}
from rgsutil import *
\end{code}
For example, the utility class {\tt OuterHull} draws the outer hull of a staircase cut from given set of 2-dimensional cut values.


\bibliography{bibliography.bib}

\begin{thebibliography}{63}
\expandafter\ifx\csname natexlab\endcsname\relax\def\natexlab#1{#1}\fi
\expandafter\ifx\csname bibnamefont\endcsname\relax
  \def\bibnamefont#1{#1}\fi
\expandafter\ifx\csname bibfnamefont\endcsname\relax
  \def\bibfnamefont#1{#1}\fi
\expandafter\ifx\csname citenamefont\endcsname\relax
  \def\citenamefont#1{#1}\fi
\expandafter\ifx\csname url\endcsname\relax
  \def\url#1{\texttt{#1}}\fi
\expandafter\ifx\csname urlprefix\endcsname\relax\def\urlprefix{URL }\fi
\providecommand{\bibinfo}[2]{#2}
\providecommand{\eprint}[2][]{\url{#2}}

\bibitem[{\citenamefont{Bhat}(2011)}]{Bhat:2010zz}
\bibinfo{author}{\bibfnamefont{P.~C.} \bibnamefont{Bhat}},
  \bibinfo{journal}{Ann. Rev. Nucl. Part. Sci.} \textbf{\bibinfo{volume}{61}},
  \bibinfo{pages}{281} (\bibinfo{year}{2011}).

\bibitem[{\citenamefont{Bishop}(2006)}]{Bishop}
\bibinfo{author}{\bibfnamefont{C.~M.} \bibnamefont{Bishop}},
  \emph{\bibinfo{title}{{Pattern Recognition and Machine Learning}}}
  (\bibinfo{publisher}{Springer}, \bibinfo{year}{2006}), ISBN
  \bibinfo{isbn}{ISBN 978-0-387-31073-2},
  \urlprefix\url{http://www.springer.com/us/book/9780387310732}.

\bibitem[{\citenamefont{Abbott et~al.}(1998)}]{Abbott:1998dc}
\bibinfo{author}{\bibfnamefont{B.}~\bibnamefont{Abbott}} \bibnamefont{et~al.}
  (\bibinfo{collaboration}{D0}), \bibinfo{journal}{Phys. Rev.}
  \textbf{\bibinfo{volume}{D58}}, \bibinfo{pages}{052001}
  (\bibinfo{year}{1998}), \eprint{hep-ex/9801025}.

\bibitem[{\citenamefont{Abazov et~al.}(2009)}]{Abazov:2009ii}
\bibinfo{author}{\bibfnamefont{V.~M.} \bibnamefont{Abazov}}
  \bibnamefont{et~al.} (\bibinfo{collaboration}{D0}), \bibinfo{journal}{Phys.
  Rev. Lett.} \textbf{\bibinfo{volume}{103}}, \bibinfo{pages}{092001}
  (\bibinfo{year}{2009}), \eprint{0903.0850}.

\bibitem[{\citenamefont{Chatrchyan
  et~al.}(2012{\natexlab{a}})}]{Chatrchyan:2012xdj}
\bibinfo{author}{\bibfnamefont{S.}~\bibnamefont{Chatrchyan}}
  \bibnamefont{et~al.} (\bibinfo{collaboration}{CMS}), \bibinfo{journal}{Phys.
  Lett.} \textbf{\bibinfo{volume}{B716}}, \bibinfo{pages}{30}
  (\bibinfo{year}{2012}{\natexlab{a}}), \eprint{1207.7235}.

\bibitem[{\citenamefont{Aad et~al.}(2012)}]{Aad:2012tfa}
\bibinfo{author}{\bibfnamefont{G.}~\bibnamefont{Aad}} \bibnamefont{et~al.}
  (\bibinfo{collaboration}{ATLAS}), \bibinfo{journal}{Phys. Lett.}
  \textbf{\bibinfo{volume}{B716}}, \bibinfo{pages}{1} (\bibinfo{year}{2012}),
  \eprint{1207.7214}.

\bibitem[{\citenamefont{Aad et~al.}(2016)}]{Aad:2015ydr}
\bibinfo{author}{\bibfnamefont{G.}~\bibnamefont{Aad}} \bibnamefont{et~al.}
  (\bibinfo{collaboration}{ATLAS}), \bibinfo{journal}{JINST}
  \textbf{\bibinfo{volume}{11}}, \bibinfo{pages}{P04008}
  (\bibinfo{year}{2016}), \eprint{1512.01094}.

\bibitem[{\citenamefont{Khachatryan et~al.}(2015)}]{Khachatryan:2015hwa}
\bibinfo{author}{\bibfnamefont{V.}~\bibnamefont{Khachatryan}}
  \bibnamefont{et~al.} (\bibinfo{collaboration}{CMS}), \bibinfo{journal}{JINST}
  \textbf{\bibinfo{volume}{10}}, \bibinfo{pages}{P06005}
  (\bibinfo{year}{2015}), \eprint{1502.02701}.

\bibitem[{\citenamefont{Sirunyan et~al.}(2017)}]{Sirunyan:2017ezt}
\bibinfo{author}{\bibfnamefont{A.~M.} \bibnamefont{Sirunyan}}
  \bibnamefont{et~al.} (\bibinfo{collaboration}{CMS}) (\bibinfo{year}{2017}),
  \eprint{1712.07158}.

\bibitem[{\citenamefont{Aaij et~al.}(2015)}]{Aaij:2015yqa}
\bibinfo{author}{\bibfnamefont{R.}~\bibnamefont{Aaij}} \bibnamefont{et~al.}
  (\bibinfo{collaboration}{LHCb}), \bibinfo{journal}{JINST}
  \textbf{\bibinfo{volume}{10}}, \bibinfo{pages}{P06013}
  (\bibinfo{year}{2015}), \eprint{1504.07670}.

\bibitem[{\citenamefont{Baldi et~al.}(2014)\citenamefont{Baldi, Sadowski, and
  Whiteson}}]{Baldi:2014kfa}
\bibinfo{author}{\bibfnamefont{P.}~\bibnamefont{Baldi}},
  \bibinfo{author}{\bibfnamefont{P.}~\bibnamefont{Sadowski}}, \bibnamefont{and}
  \bibinfo{author}{\bibfnamefont{D.}~\bibnamefont{Whiteson}},
  \bibinfo{journal}{Nature Commun.} \textbf{\bibinfo{volume}{5}},
  \bibinfo{pages}{4308} (\bibinfo{year}{2014}), \eprint{1402.4735}.

\bibitem[{\citenamefont{Baldi et~al.}(2015)\citenamefont{Baldi, Sadowski, and
  Whiteson}}]{Baldi:2014pta}
\bibinfo{author}{\bibfnamefont{P.}~\bibnamefont{Baldi}},
  \bibinfo{author}{\bibfnamefont{P.}~\bibnamefont{Sadowski}}, \bibnamefont{and}
  \bibinfo{author}{\bibfnamefont{D.}~\bibnamefont{Whiteson}},
  \bibinfo{journal}{Phys. Rev. Lett.} \textbf{\bibinfo{volume}{114}},
  \bibinfo{pages}{111801} (\bibinfo{year}{2015}), \eprint{1410.3469}.

\bibitem[{\citenamefont{Baldi et~al.}(2016)\citenamefont{Baldi, Bauer, Eng,
  Sadowski, and Whiteson}}]{Baldi:2016fql}
\bibinfo{author}{\bibfnamefont{P.}~\bibnamefont{Baldi}},
  \bibinfo{author}{\bibfnamefont{K.}~\bibnamefont{Bauer}},
  \bibinfo{author}{\bibfnamefont{C.}~\bibnamefont{Eng}},
  \bibinfo{author}{\bibfnamefont{P.}~\bibnamefont{Sadowski}}, \bibnamefont{and}
  \bibinfo{author}{\bibfnamefont{D.}~\bibnamefont{Whiteson}},
  \bibinfo{journal}{Phys. Rev.} \textbf{\bibinfo{volume}{D93}},
  \bibinfo{pages}{094034} (\bibinfo{year}{2016}), \eprint{1603.09349}.

\bibitem[{\citenamefont{Guest et~al.}(2016)\citenamefont{Guest, Collado, Baldi,
  Hsu, Urban, and Whiteson}}]{Guest:2016iqz}
\bibinfo{author}{\bibfnamefont{D.}~\bibnamefont{Guest}},
  \bibinfo{author}{\bibfnamefont{J.}~\bibnamefont{Collado}},
  \bibinfo{author}{\bibfnamefont{P.}~\bibnamefont{Baldi}},
  \bibinfo{author}{\bibfnamefont{S.-C.} \bibnamefont{Hsu}},
  \bibinfo{author}{\bibfnamefont{G.}~\bibnamefont{Urban}}, \bibnamefont{and}
  \bibinfo{author}{\bibfnamefont{D.}~\bibnamefont{Whiteson}},
  \bibinfo{journal}{Phys. Rev.} \textbf{\bibinfo{volume}{D94}},
  \bibinfo{pages}{112002} (\bibinfo{year}{2016}), \eprint{1607.08633}.

\bibitem[{\citenamefont{Searcy et~al.}(2016)\citenamefont{Searcy, Huang,
  Pleier, and Zhu}}]{Searcy:2015apa}
\bibinfo{author}{\bibfnamefont{J.}~\bibnamefont{Searcy}},
  \bibinfo{author}{\bibfnamefont{L.}~\bibnamefont{Huang}},
  \bibinfo{author}{\bibfnamefont{M.-A.} \bibnamefont{Pleier}},
  \bibnamefont{and} \bibinfo{author}{\bibfnamefont{J.}~\bibnamefont{Zhu}},
  \bibinfo{journal}{Phys. Rev.} \textbf{\bibinfo{volume}{D93}},
  \bibinfo{pages}{094033} (\bibinfo{year}{2016}), \eprint{1510.01691}.

\bibitem[{\citenamefont{Aurisano et~al.}(2016)\citenamefont{Aurisano, Radovic,
  Rocco, Himmel, Messier, Niner, Pawloski, Psihas, Sousa, and
  Vahle}}]{Aurisano:2016jvx}
\bibinfo{author}{\bibfnamefont{A.}~\bibnamefont{Aurisano}},
  \bibinfo{author}{\bibfnamefont{A.}~\bibnamefont{Radovic}},
  \bibinfo{author}{\bibfnamefont{D.}~\bibnamefont{Rocco}},
  \bibinfo{author}{\bibfnamefont{A.}~\bibnamefont{Himmel}},
  \bibinfo{author}{\bibfnamefont{M.~D.} \bibnamefont{Messier}},
  \bibinfo{author}{\bibfnamefont{E.}~\bibnamefont{Niner}},
  \bibinfo{author}{\bibfnamefont{G.}~\bibnamefont{Pawloski}},
  \bibinfo{author}{\bibfnamefont{F.}~\bibnamefont{Psihas}},
  \bibinfo{author}{\bibfnamefont{A.}~\bibnamefont{Sousa}}, \bibnamefont{and}
  \bibinfo{author}{\bibfnamefont{P.}~\bibnamefont{Vahle}},
  \bibinfo{journal}{JINST} \textbf{\bibinfo{volume}{11}},
  \bibinfo{pages}{P09001} (\bibinfo{year}{2016}), \eprint{1604.01444}.

\bibitem[{\citenamefont{Acciarri et~al.}(2017)}]{Acciarri:2016ryt}
\bibinfo{author}{\bibfnamefont{R.}~\bibnamefont{Acciarri}} \bibnamefont{et~al.}
  (\bibinfo{collaboration}{MicroBooNE}), \bibinfo{journal}{JINST}
  \textbf{\bibinfo{volume}{12}}, \bibinfo{pages}{P03011}
  (\bibinfo{year}{2017}), \eprint{1611.05531}.

\bibitem[{\citenamefont{Hoecker et~al.}(2007)\citenamefont{Hoecker, Speckmayer,
  Stelzer, Therhaag, von Toerne, and Voss}}]{Hocker:2007ht}
\bibinfo{author}{\bibfnamefont{A.}~\bibnamefont{Hoecker}},
  \bibinfo{author}{\bibfnamefont{P.}~\bibnamefont{Speckmayer}},
  \bibinfo{author}{\bibfnamefont{J.}~\bibnamefont{Stelzer}},
  \bibinfo{author}{\bibfnamefont{J.}~\bibnamefont{Therhaag}},
  \bibinfo{author}{\bibfnamefont{E.}~\bibnamefont{von Toerne}},
  \bibnamefont{and} \bibinfo{author}{\bibfnamefont{H.}~\bibnamefont{Voss}},
  \bibinfo{journal}{PoS} \textbf{\bibinfo{volume}{ACAT}}, \bibinfo{pages}{040}
  (\bibinfo{year}{2007}), \eprint{physics/0703039}.

\bibitem[{\citenamefont{Feindt and Kerzel}(2006)}]{Feindt:2006pm}
\bibinfo{author}{\bibfnamefont{M.}~\bibnamefont{Feindt}} \bibnamefont{and}
  \bibinfo{author}{\bibfnamefont{U.}~\bibnamefont{Kerzel}},
  \bibinfo{journal}{Nucl. Instrum. Meth.} \textbf{\bibinfo{volume}{A559}},
  \bibinfo{pages}{190} (\bibinfo{year}{2006}).

\bibitem[{\citenamefont{Pedregosa et~al.}(2011)\citenamefont{Pedregosa,
  Varoquaux, Gramfort, Michel, Thirion, Grisel, Blondel, Prettenhofer, Weiss,
  Dubourg et~al.}}]{scikit-learn}
\bibinfo{author}{\bibfnamefont{F.}~\bibnamefont{Pedregosa}},
  \bibinfo{author}{\bibfnamefont{G.}~\bibnamefont{Varoquaux}},
  \bibinfo{author}{\bibfnamefont{A.}~\bibnamefont{Gramfort}},
  \bibinfo{author}{\bibfnamefont{V.}~\bibnamefont{Michel}},
  \bibinfo{author}{\bibfnamefont{B.}~\bibnamefont{Thirion}},
  \bibinfo{author}{\bibfnamefont{O.}~\bibnamefont{Grisel}},
  \bibinfo{author}{\bibfnamefont{M.}~\bibnamefont{Blondel}},
  \bibinfo{author}{\bibfnamefont{P.}~\bibnamefont{Prettenhofer}},
  \bibinfo{author}{\bibfnamefont{R.}~\bibnamefont{Weiss}},
  \bibinfo{author}{\bibfnamefont{V.}~\bibnamefont{Dubourg}},
  \bibnamefont{et~al.}, \bibinfo{journal}{Journal of Machine Learning Research}
  \textbf{\bibinfo{volume}{12}}, \bibinfo{pages}{2825} (\bibinfo{year}{2011}).

\bibitem[{\citenamefont{Chollet et~al.}(2015)}]{chollet2015keras}
\bibinfo{author}{\bibfnamefont{F.}~\bibnamefont{Chollet}} \bibnamefont{et~al.},
  \emph{\bibinfo{title}{Keras}},
  \bibinfo{howpublished}{\url{https://github.com/keras-team/keras}}
  (\bibinfo{year}{2015}).

\bibitem[{\citenamefont{Abadi et~al.}(2015)\citenamefont{Abadi, Agarwal,
  Barham, Brevdo, Chen, Citro, Corrado, Davis, Dean, Devin
  et~al.}}]{tensorflow2015-whitepaper}
\bibinfo{author}{\bibfnamefont{M.}~\bibnamefont{Abadi}},
  \bibinfo{author}{\bibfnamefont{A.}~\bibnamefont{Agarwal}},
  \bibinfo{author}{\bibfnamefont{P.}~\bibnamefont{Barham}},
  \bibinfo{author}{\bibfnamefont{E.}~\bibnamefont{Brevdo}},
  \bibinfo{author}{\bibfnamefont{Z.}~\bibnamefont{Chen}},
  \bibinfo{author}{\bibfnamefont{C.}~\bibnamefont{Citro}},
  \bibinfo{author}{\bibfnamefont{G.~S.} \bibnamefont{Corrado}},
  \bibinfo{author}{\bibfnamefont{A.}~\bibnamefont{Davis}},
  \bibinfo{author}{\bibfnamefont{J.}~\bibnamefont{Dean}},
  \bibinfo{author}{\bibfnamefont{M.}~\bibnamefont{Devin}},
  \bibnamefont{et~al.}, \emph{\bibinfo{title}{{TensorFlow}: Large-scale machine
  learning on heterogeneous systems}} (\bibinfo{year}{2015}),
  \bibinfo{note}{software available from tensorflow.org},
  \urlprefix\url{https://www.tensorflow.org/}.

\bibitem[{\citenamefont{Abadi et~al.}(2016)\citenamefont{Abadi, Barham, Chen,
  Chen, Davis, Dean, Devin, Ghemawat, Irving, Isard et~al.}}]{45381}
\bibinfo{author}{\bibfnamefont{M.}~\bibnamefont{Abadi}},
  \bibinfo{author}{\bibfnamefont{P.}~\bibnamefont{Barham}},
  \bibinfo{author}{\bibfnamefont{J.}~\bibnamefont{Chen}},
  \bibinfo{author}{\bibfnamefont{Z.}~\bibnamefont{Chen}},
  \bibinfo{author}{\bibfnamefont{A.}~\bibnamefont{Davis}},
  \bibinfo{author}{\bibfnamefont{J.}~\bibnamefont{Dean}},
  \bibinfo{author}{\bibfnamefont{M.}~\bibnamefont{Devin}},
  \bibinfo{author}{\bibfnamefont{S.}~\bibnamefont{Ghemawat}},
  \bibinfo{author}{\bibfnamefont{G.}~\bibnamefont{Irving}},
  \bibinfo{author}{\bibfnamefont{M.}~\bibnamefont{Isard}},
  \bibnamefont{et~al.}, in \emph{\bibinfo{booktitle}{12th USENIX Symposium on
  Operating Systems Design and Implementation (OSDI 16)}}
  (\bibinfo{year}{2016}), pp. \bibinfo{pages}{265--283},
  \urlprefix\url{https://www.usenix.org/system/files/conference/osdi16/osdi16-abadi.pdf}.

\bibitem[{\citenamefont{{Theano Development
  Team}}(2016)}]{2016arXiv160502688short}
\bibinfo{author}{\bibnamefont{{Theano Development Team}}},
  \bibinfo{journal}{arXiv e-prints} \textbf{\bibinfo{volume}{abs/1605.02688}}
  (\bibinfo{year}{2016}), \urlprefix\url{http://arxiv.org/abs/1605.02688}.

\bibitem[{\citenamefont{Ballintijn et~al.}(2006)\citenamefont{Ballintijn,
  Biskup, Brun, Ganis, Kickinger, Peters, Rademakers, Canal, and
  Feichtinger}}]{Ballintijn:2006ni}
\bibinfo{author}{\bibfnamefont{M.}~\bibnamefont{Ballintijn}},
  \bibinfo{author}{\bibfnamefont{M.}~\bibnamefont{Biskup}},
  \bibinfo{author}{\bibfnamefont{R.}~\bibnamefont{Brun}},
  \bibinfo{author}{\bibfnamefont{G.}~\bibnamefont{Ganis}},
  \bibinfo{author}{\bibfnamefont{G.}~\bibnamefont{Kickinger}},
  \bibinfo{author}{\bibfnamefont{A.}~\bibnamefont{Peters}},
  \bibinfo{author}{\bibfnamefont{F.}~\bibnamefont{Rademakers}},
  \bibinfo{author}{\bibfnamefont{P.}~\bibnamefont{Canal}}, \bibnamefont{and}
  \bibinfo{author}{\bibfnamefont{D.}~\bibnamefont{Feichtinger}},
  \bibinfo{journal}{Nucl. Instrum. Meth.} \textbf{\bibinfo{volume}{A559}},
  \bibinfo{pages}{13} (\bibinfo{year}{2006}).

\bibitem[{\citenamefont{Bhat et~al.}(1993-1994)\citenamefont{Bhat, Prosper, and
  Stewart}}]{rgs0}
\bibinfo{author}{\bibfnamefont{P.~C.} \bibnamefont{Bhat}},
  \bibinfo{author}{\bibfnamefont{H.~B.} \bibnamefont{Prosper}},
  \bibnamefont{and} \bibinfo{author}{\bibfnamefont{C.}~\bibnamefont{Stewart}}
  (\bibinfo{year}{1993-1994}), \bibinfo{note}{unpublished}.

\bibitem[{\citenamefont{Amos et~al.}(1995)\citenamefont{Amos, Stewart, Bhat,
  Cretsinger, Won, Dharmaratna, and Prosper}}]{Amos:1995tn}
\bibinfo{author}{\bibfnamefont{N.~A.} \bibnamefont{Amos}},
  \bibinfo{author}{\bibfnamefont{C.}~\bibnamefont{Stewart}},
  \bibinfo{author}{\bibfnamefont{P.}~\bibnamefont{Bhat}},
  \bibinfo{author}{\bibfnamefont{C.}~\bibnamefont{Cretsinger}},
  \bibinfo{author}{\bibfnamefont{E.}~\bibnamefont{Won}},
  \bibinfo{author}{\bibfnamefont{W.~G.~D.} \bibnamefont{Dharmaratna}},
  \bibnamefont{and} \bibinfo{author}{\bibfnamefont{H.~B.}
  \bibnamefont{Prosper}}, in \emph{\bibinfo{booktitle}{{Proceedings, 8th
  International Conference on Computing in High-Energy and Nuclear Physics
  (CHEP 1995): Rio de Janeiro, Brazil, September 18-22, 1995}}}
  (\bibinfo{year}{1995}), pp. \bibinfo{pages}{215--219}.

\bibitem[{\citenamefont{Abachi et~al.}(1995)}]{Abachi:1995iq}
\bibinfo{author}{\bibfnamefont{S.}~\bibnamefont{Abachi}} \bibnamefont{et~al.}
  (\bibinfo{collaboration}{D0}), \bibinfo{journal}{Phys. Rev. Lett.}
  \textbf{\bibinfo{volume}{74}}, \bibinfo{pages}{2632} (\bibinfo{year}{1995}),
  \eprint{hep-ex/9503003}.

\bibitem[{\citenamefont{Abe et~al.}(1995)}]{Abe:1995hr}
\bibinfo{author}{\bibfnamefont{F.}~\bibnamefont{Abe}} \bibnamefont{et~al.}
  (\bibinfo{collaboration}{CDF}), \bibinfo{journal}{Phys. Rev. Lett.}
  \textbf{\bibinfo{volume}{74}}, \bibinfo{pages}{2626} (\bibinfo{year}{1995}),
  \eprint{hep-ex/9503002}.

\bibitem[{\citenamefont{Abazov et~al.}(2001)}]{Abazov:2001mx}
\bibinfo{author}{\bibfnamefont{V.~M.} \bibnamefont{Abazov}}
  \bibnamefont{et~al.} (\bibinfo{collaboration}{D0}), \bibinfo{journal}{Phys.
  Rev.} \textbf{\bibinfo{volume}{D64}}, \bibinfo{pages}{092004}
  (\bibinfo{year}{2001}), \eprint{hep-ex/0105072}.

\bibitem[{\citenamefont{Abazov et~al.}(2003)}]{Abazov:2002gy}
\bibinfo{author}{\bibfnamefont{V.~M.} \bibnamefont{Abazov}}
  \bibnamefont{et~al.} (\bibinfo{collaboration}{D0}), \bibinfo{journal}{Phys.
  Rev.} \textbf{\bibinfo{volume}{D67}}, \bibinfo{pages}{012004}
  (\bibinfo{year}{2003}), \eprint{hep-ex/0205019}.

\bibitem[{\citenamefont{Chatrchyan
  et~al.}(2012{\natexlab{b}})}]{Chatrchyan:2012rga}
\bibinfo{author}{\bibfnamefont{S.}~\bibnamefont{Chatrchyan}}
  \bibnamefont{et~al.} (\bibinfo{collaboration}{CMS}), \bibinfo{journal}{JHEP}
  \textbf{\bibinfo{volume}{04}}, \bibinfo{pages}{033}
  (\bibinfo{year}{2012}{\natexlab{b}}), \eprint{1203.3976}.

\bibitem[{\citenamefont{Strobbe}(2011)}]{Strobbe:2011lta}
\bibinfo{author}{\bibfnamefont{N.~C.} \bibnamefont{Strobbe}}, Master's thesis,
  \bibinfo{school}{Gent U.} (\bibinfo{year}{2011}),
  \urlprefix\url{http://inspirehep.net/record/1088189/files/openfile.pdf}.

\bibitem[{\citenamefont{Rogan}(2010)}]{Rogan:2010kb}
\bibinfo{author}{\bibfnamefont{C.}~\bibnamefont{Rogan}} (\bibinfo{year}{2010}),
  \eprint{1006.2727}.

\bibitem[{\citenamefont{Patrignani et~al.}(2016)}]{Olive:2016xmw}
\bibinfo{author}{\bibfnamefont{C.}~\bibnamefont{Patrignani}}
  \bibnamefont{et~al.} (\bibinfo{collaboration}{Particle Data Group}),
  \bibinfo{journal}{Chin. Phys.} \textbf{\bibinfo{volume}{C40}},
  \bibinfo{pages}{100001} (\bibinfo{year}{2016}).

\bibitem[{\citenamefont{Kullback and Leibler}(1951)}]{KL}
\bibinfo{author}{\bibfnamefont{S.}~\bibnamefont{Kullback}} \bibnamefont{and}
  \bibinfo{author}{\bibfnamefont{R.~A.} \bibnamefont{Leibler}},
  \bibinfo{journal}{Ann. Math. Statist.} \textbf{\bibinfo{volume}{22 (1)}},
  \bibinfo{pages}{79} (\bibinfo{year}{1951}).

\bibitem[{\citenamefont{Collaboration}(2016)}]{CMS:2016jjx}
\bibinfo{author}{\bibfnamefont{C.}~\bibnamefont{Collaboration}}
  (\bibinfo{collaboration}{CMS}) (\bibinfo{year}{2016}).

\bibitem[{\citenamefont{collaboration}(2016)}]{ATLAS:2016gld}
\bibinfo{author}{\bibfnamefont{T.~A.} \bibnamefont{collaboration}}
  (\bibinfo{collaboration}{ATLAS}) (\bibinfo{year}{2016}).

\bibitem[{\citenamefont{Aaboud et~al.}(2016)}]{Aaboud:2016cns}
\bibinfo{author}{\bibfnamefont{M.}~\bibnamefont{Aaboud}} \bibnamefont{et~al.}
  (\bibinfo{collaboration}{ATLAS}), \bibinfo{journal}{JHEP}
  \textbf{\bibinfo{volume}{11}}, \bibinfo{pages}{112} (\bibinfo{year}{2016}),
  \eprint{1606.02181}.

\bibitem[{\citenamefont{Cacciari et~al.}(2015)\citenamefont{Cacciari, Dreyer,
  Karlberg, Salam, and Zanderighi}}]{Cacciari:2015jma}
\bibinfo{author}{\bibfnamefont{M.}~\bibnamefont{Cacciari}},
  \bibinfo{author}{\bibfnamefont{F.~A.} \bibnamefont{Dreyer}},
  \bibinfo{author}{\bibfnamefont{A.}~\bibnamefont{Karlberg}},
  \bibinfo{author}{\bibfnamefont{G.~P.} \bibnamefont{Salam}}, \bibnamefont{and}
  \bibinfo{author}{\bibfnamefont{G.}~\bibnamefont{Zanderighi}},
  \bibinfo{journal}{Phys. Rev. Lett.} \textbf{\bibinfo{volume}{115}},
  \bibinfo{pages}{082002} (\bibinfo{year}{2015}), \eprint{1506.02660}.

\bibitem[{\citenamefont{Chatrchyan et~al.}(2014)}]{Chatrchyan:2013mxa}
\bibinfo{author}{\bibfnamefont{S.}~\bibnamefont{Chatrchyan}}
  \bibnamefont{et~al.} (\bibinfo{collaboration}{CMS}), \bibinfo{journal}{Phys.
  Rev.} \textbf{\bibinfo{volume}{D89}}, \bibinfo{pages}{092007}
  (\bibinfo{year}{2014}), \eprint{1312.5353}.

\bibitem[{\citenamefont{Sjostrand et~al.}(2006)\citenamefont{Sjostrand, Mrenna,
  and Skands}}]{Sjostrand:2006za}
\bibinfo{author}{\bibfnamefont{T.}~\bibnamefont{Sjostrand}},
  \bibinfo{author}{\bibfnamefont{S.}~\bibnamefont{Mrenna}}, \bibnamefont{and}
  \bibinfo{author}{\bibfnamefont{P.~Z.} \bibnamefont{Skands}},
  \bibinfo{journal}{JHEP} \textbf{\bibinfo{volume}{05}}, \bibinfo{pages}{026}
  (\bibinfo{year}{2006}), \eprint{hep-ph/0603175}.

\bibitem[{\citenamefont{Sjostrand et~al.}(2008)\citenamefont{Sjostrand, Mrenna,
  and Skands}}]{Sjostrand:2007gs}
\bibinfo{author}{\bibfnamefont{T.}~\bibnamefont{Sjostrand}},
  \bibinfo{author}{\bibfnamefont{S.}~\bibnamefont{Mrenna}}, \bibnamefont{and}
  \bibinfo{author}{\bibfnamefont{P.~Z.} \bibnamefont{Skands}},
  \bibinfo{journal}{Comput. Phys. Commun.} \textbf{\bibinfo{volume}{178}},
  \bibinfo{pages}{852} (\bibinfo{year}{2008}), \eprint{0710.3820}.

\bibitem[{\citenamefont{de~Favereau et~al.}(2014)\citenamefont{de~Favereau,
  Delaere, Demin, Giammanco, Lemaître, Mertens, and
  Selvaggi}}]{deFavereau:2013fsa}
\bibinfo{author}{\bibfnamefont{J.}~\bibnamefont{de~Favereau}},
  \bibinfo{author}{\bibfnamefont{C.}~\bibnamefont{Delaere}},
  \bibinfo{author}{\bibfnamefont{P.}~\bibnamefont{Demin}},
  \bibinfo{author}{\bibfnamefont{A.}~\bibnamefont{Giammanco}},
  \bibinfo{author}{\bibfnamefont{V.}~\bibnamefont{Lemaître}},
  \bibinfo{author}{\bibfnamefont{A.}~\bibnamefont{Mertens}}, \bibnamefont{and}
  \bibinfo{author}{\bibfnamefont{M.}~\bibnamefont{Selvaggi}}
  (\bibinfo{collaboration}{DELPHES 3}), \bibinfo{journal}{JHEP}
  \textbf{\bibinfo{volume}{02}}, \bibinfo{pages}{057} (\bibinfo{year}{2014}),
  \eprint{1307.6346}.

\bibitem[{\citenamefont{Chatrchyan et~al.}(2013)}]{Chatrchyan:2013sba}
\bibinfo{author}{\bibfnamefont{S.}~\bibnamefont{Chatrchyan}}
  \bibnamefont{et~al.} (\bibinfo{collaboration}{CMS}), \bibinfo{journal}{JINST}
  \textbf{\bibinfo{volume}{8}}, \bibinfo{pages}{P11002} (\bibinfo{year}{2013}),
  \eprint{1306.6905}.

\bibitem[{\citenamefont{Nason and Zanderighi}(2014)}]{Nason:2013ydw}
\bibinfo{author}{\bibfnamefont{P.}~\bibnamefont{Nason}} \bibnamefont{and}
  \bibinfo{author}{\bibfnamefont{G.}~\bibnamefont{Zanderighi}},
  \bibinfo{journal}{Eur. Phys. J.} \textbf{\bibinfo{volume}{C74}},
  \bibinfo{pages}{2702} (\bibinfo{year}{2014}), \eprint{1311.1365}.

\bibitem[{\citenamefont{Dulat et~al.}(2016)\citenamefont{Dulat, Hou, Gao,
  Guzzi, Huston, Nadolsky, Pumplin, Schmidt, Stump, and Yuan}}]{Dulat:2015mca}
\bibinfo{author}{\bibfnamefont{S.}~\bibnamefont{Dulat}},
  \bibinfo{author}{\bibfnamefont{T.-J.} \bibnamefont{Hou}},
  \bibinfo{author}{\bibfnamefont{J.}~\bibnamefont{Gao}},
  \bibinfo{author}{\bibfnamefont{M.}~\bibnamefont{Guzzi}},
  \bibinfo{author}{\bibfnamefont{J.}~\bibnamefont{Huston}},
  \bibinfo{author}{\bibfnamefont{P.}~\bibnamefont{Nadolsky}},
  \bibinfo{author}{\bibfnamefont{J.}~\bibnamefont{Pumplin}},
  \bibinfo{author}{\bibfnamefont{C.}~\bibnamefont{Schmidt}},
  \bibinfo{author}{\bibfnamefont{D.}~\bibnamefont{Stump}}, \bibnamefont{and}
  \bibinfo{author}{\bibfnamefont{C.~P.} \bibnamefont{Yuan}},
  \bibinfo{journal}{Phys. Rev.} \textbf{\bibinfo{volume}{D93}},
  \bibinfo{pages}{033006} (\bibinfo{year}{2016}), \eprint{1506.07443}.

\bibitem[{\citenamefont{Buckley et~al.}(2015)\citenamefont{Buckley, Ferrando,
  Lloyd, Nordström, Page, Rüfenacht, Schönherr, and Watt}}]{Buckley:2014ana}
\bibinfo{author}{\bibfnamefont{A.}~\bibnamefont{Buckley}},
  \bibinfo{author}{\bibfnamefont{J.}~\bibnamefont{Ferrando}},
  \bibinfo{author}{\bibfnamefont{S.}~\bibnamefont{Lloyd}},
  \bibinfo{author}{\bibfnamefont{K.}~\bibnamefont{Nordström}},
  \bibinfo{author}{\bibfnamefont{B.}~\bibnamefont{Page}},
  \bibinfo{author}{\bibfnamefont{M.}~\bibnamefont{Rüfenacht}},
  \bibinfo{author}{\bibfnamefont{M.}~\bibnamefont{Schönherr}},
  \bibnamefont{and} \bibinfo{author}{\bibfnamefont{G.}~\bibnamefont{Watt}},
  \bibinfo{journal}{Eur. Phys. J.} \textbf{\bibinfo{volume}{C75}},
  \bibinfo{pages}{132} (\bibinfo{year}{2015}), \eprint{1412.7420}.

\bibitem[{\citenamefont{Campbell et~al.}(2015)\citenamefont{Campbell, Ellis,
  and Giele}}]{Campbell:2015qma}
\bibinfo{author}{\bibfnamefont{J.~M.} \bibnamefont{Campbell}},
  \bibinfo{author}{\bibfnamefont{R.~K.} \bibnamefont{Ellis}}, \bibnamefont{and}
  \bibinfo{author}{\bibfnamefont{W.~T.} \bibnamefont{Giele}},
  \bibinfo{journal}{Eur. Phys. J.} \textbf{\bibinfo{volume}{C75}},
  \bibinfo{pages}{246} (\bibinfo{year}{2015}), \eprint{1503.06182}.

\bibitem[{\citenamefont{de~Florian et~al.}(2016)}]{deFlorian:2016spz}
\bibinfo{author}{\bibfnamefont{D.}~\bibnamefont{de~Florian}}
  \bibnamefont{et~al.} (\bibinfo{collaboration}{LHC Higgs Cross Section Working
  Group}) (\bibinfo{year}{2016}), \eprint{1610.07922}.

\bibitem[{\citenamefont{Aad et~al.}(2015)}]{Aad:2014eva}
\bibinfo{author}{\bibfnamefont{G.}~\bibnamefont{Aad}} \bibnamefont{et~al.}
  (\bibinfo{collaboration}{ATLAS}), \bibinfo{journal}{Phys. Rev.}
  \textbf{\bibinfo{volume}{D91}}, \bibinfo{pages}{012006}
  (\bibinfo{year}{2015}), \eprint{1408.5191}.

\bibitem[{\citenamefont{Khachatryan et~al.}(2016)}]{Khachatryan:2016zcu}
\bibinfo{author}{\bibfnamefont{V.}~\bibnamefont{Khachatryan}}
  \bibnamefont{et~al.} (\bibinfo{collaboration}{CMS}), \bibinfo{journal}{Phys.
  Rev.} \textbf{\bibinfo{volume}{D93}}, \bibinfo{pages}{092009}
  (\bibinfo{year}{2016}), \eprint{1602.02917}.

\bibitem[{\citenamefont{Allanach}(2002)}]{Allanach:2001kg}
\bibinfo{author}{\bibfnamefont{B.~C.} \bibnamefont{Allanach}},
  \bibinfo{journal}{Comput. Phys. Commun.} \textbf{\bibinfo{volume}{143}},
  \bibinfo{pages}{305} (\bibinfo{year}{2002}), \eprint{hep-ph/0104145}.

\bibitem[{\citenamefont{Djouadi et~al.}(2007)\citenamefont{Djouadi,
  Muhlleitner, and Spira}}]{Djouadi:2006bz}
\bibinfo{author}{\bibfnamefont{A.}~\bibnamefont{Djouadi}},
  \bibinfo{author}{\bibfnamefont{M.~M.} \bibnamefont{Muhlleitner}},
  \bibnamefont{and} \bibinfo{author}{\bibfnamefont{M.}~\bibnamefont{Spira}},
  \bibinfo{journal}{Acta Phys. Polon.} \textbf{\bibinfo{volume}{B38}},
  \bibinfo{pages}{635} (\bibinfo{year}{2007}), \eprint{hep-ph/0609292}.

\bibitem[{\citenamefont{Beenakker et~al.}(1996)\citenamefont{Beenakker, Hopker,
  and Spira}}]{Beenakker:1996ed}
\bibinfo{author}{\bibfnamefont{W.}~\bibnamefont{Beenakker}},
  \bibinfo{author}{\bibfnamefont{R.}~\bibnamefont{Hopker}}, \bibnamefont{and}
  \bibinfo{author}{\bibfnamefont{M.}~\bibnamefont{Spira}}
  (\bibinfo{year}{1996}), \eprint{hep-ph/9611232}.

\bibitem[{\citenamefont{Beenakker et~al.}(1997)\citenamefont{Beenakker, Hopker,
  Spira, and Zerwas}}]{Beenakker:1996ch}
\bibinfo{author}{\bibfnamefont{W.}~\bibnamefont{Beenakker}},
  \bibinfo{author}{\bibfnamefont{R.}~\bibnamefont{Hopker}},
  \bibinfo{author}{\bibfnamefont{M.}~\bibnamefont{Spira}}, \bibnamefont{and}
  \bibinfo{author}{\bibfnamefont{P.~M.} \bibnamefont{Zerwas}},
  \bibinfo{journal}{Nucl. Phys.} \textbf{\bibinfo{volume}{B492}},
  \bibinfo{pages}{51} (\bibinfo{year}{1997}), \eprint{hep-ph/9610490}.

\bibitem[{\citenamefont{Czakon and Mitov}(2014)}]{Czakon:2011xx}
\bibinfo{author}{\bibfnamefont{M.}~\bibnamefont{Czakon}} \bibnamefont{and}
  \bibinfo{author}{\bibfnamefont{A.}~\bibnamefont{Mitov}},
  \bibinfo{journal}{Comput. Phys. Commun.} \textbf{\bibinfo{volume}{185}},
  \bibinfo{pages}{2930} (\bibinfo{year}{2014}), \eprint{1112.5675}.

\bibitem[{\citenamefont{Czakon et~al.}(2013)\citenamefont{Czakon, Fiedler, and
  Mitov}}]{Czakon:2013goa}
\bibinfo{author}{\bibfnamefont{M.}~\bibnamefont{Czakon}},
  \bibinfo{author}{\bibfnamefont{P.}~\bibnamefont{Fiedler}}, \bibnamefont{and}
  \bibinfo{author}{\bibfnamefont{A.}~\bibnamefont{Mitov}},
  \bibinfo{journal}{Phys. Rev. Lett.} \textbf{\bibinfo{volume}{110}},
  \bibinfo{pages}{252004} (\bibinfo{year}{2013}), \eprint{1303.6254}.

\bibitem[{\citenamefont{Czakon and Mitov}(2013)}]{Czakon:2012pz}
\bibinfo{author}{\bibfnamefont{M.}~\bibnamefont{Czakon}} \bibnamefont{and}
  \bibinfo{author}{\bibfnamefont{A.}~\bibnamefont{Mitov}},
  \bibinfo{journal}{JHEP} \textbf{\bibinfo{volume}{01}}, \bibinfo{pages}{080}
  (\bibinfo{year}{2013}), \eprint{1210.6832}.

\bibitem[{\citenamefont{Czakon and Mitov}(2012)}]{Czakon:2012zr}
\bibinfo{author}{\bibfnamefont{M.}~\bibnamefont{Czakon}} \bibnamefont{and}
  \bibinfo{author}{\bibfnamefont{A.}~\bibnamefont{Mitov}},
  \bibinfo{journal}{JHEP} \textbf{\bibinfo{volume}{12}}, \bibinfo{pages}{054}
  (\bibinfo{year}{2012}), \eprint{1207.0236}.

\bibitem[{\citenamefont{Bärnreuther et~al.}(2012)\citenamefont{Bärnreuther,
  Czakon, and Mitov}}]{Baernreuther:2012ws}
\bibinfo{author}{\bibfnamefont{P.}~\bibnamefont{Bärnreuther}},
  \bibinfo{author}{\bibfnamefont{M.}~\bibnamefont{Czakon}}, \bibnamefont{and}
  \bibinfo{author}{\bibfnamefont{A.}~\bibnamefont{Mitov}},
  \bibinfo{journal}{Phys. Rev. Lett.} \textbf{\bibinfo{volume}{109}},
  \bibinfo{pages}{132001} (\bibinfo{year}{2012}), \eprint{1204.5201}.

\bibitem[{\citenamefont{{Prosper, H. B. and Sekmen, S.}}(2012)}]{tnm}
\bibinfo{author}{\bibnamefont{{Prosper, H. B. and Sekmen, S.}}},
  \bibinfo{type}{CMS Internal Note} \bibinfo{number}{CMS-IN-2012-012},
  \bibinfo{institution}{CERN} (\bibinfo{year}{2012}),
  \urlprefix\url{http://cdsweb.cern.ch/record/1279362}.

\bibitem[{\citenamefont{Thaler and Van~Tilburg}(2011)}]{Thaler:2010tr}
\bibinfo{author}{\bibfnamefont{J.}~\bibnamefont{Thaler}} \bibnamefont{and}
  \bibinfo{author}{\bibfnamefont{K.}~\bibnamefont{Van~Tilburg}},
  \bibinfo{journal}{JHEP} \textbf{\bibinfo{volume}{03}}, \bibinfo{pages}{015}
  (\bibinfo{year}{2011}), \eprint{1011.2268}.

\end{thebibliography}

\end{document}